\documentclass[11pt]{article}
\usepackage[utf8]{inputenc}
\usepackage{amsmath}
\usepackage{amssymb}
\usepackage{enumerate}
\usepackage{amsfonts}
\usepackage{nicefrac}
\usepackage{microtype}
\usepackage[margin=0.7in]{geometry}
\usepackage{graphicx}
\usepackage{setspace}
\usepackage{bbm}
\usepackage{mathtools}
\usepackage[round,numbers,sort&compress]{natbib}
\setcitestyle{square}
\usepackage{sectsty}
\usepackage{verbatim}
\usepackage[hidelinks]{hyperref}

\hypersetup{
pdftitle={An Introduction To Langer's Theory},
pdfauthor={Lu Hong}
}

\begin{document}

\title{An Introduction to Langer's Theory}

\author{Lu Hong \\
        Department of Bioengineering \& Therapeutic Sciences \\
        University of California, San Fransicso \\
        San Francisco, CA 94158 \\
        \texttt{lu.hong@ucsf.edu}}
\date{}

\maketitle

\begin{abstract}

This note provides a pedagogical introduction to Langer's theory for activated rate processes in multiple dimensions at the high friction limit, with an emphasis on the connection between the theory and the property of the backward committor/splitting probability near the saddle point. The intended audience is assumed to have some familiarity with linear algebra and statistical mechanics while knowledge of stochastic processes is not strictly necessary. \bigskip

\end{abstract}

\tableofcontents
\newpage

This note is intended as an introduction to Langer's theory \citep{langerStatisticalTheoryDecay1969}, the multidimensional extension of Kramers' theory for activated rate processes in the high friction limit \citep{kramers_brownian_1940}. The note is organized as follows: Section \ref{review_section} briefly reviews some elements of the theory of stochastic process, which prepares for the introduction of Kramers' theory in Section \ref{Kramers_theory}, where the flux-over-population method is demonstrated in the relatively simple context of a one-dimensional double-well potential. The majority of this note is devoted to the discussion of Langer's theory in Section \ref{Langer_theory}. The derivation begins by relating the steady-state probability density in the flux-over-population method to the backward committor, followed by a detailed analysis of the behavior of the committor near the saddle point. This allows us to derive an expression for the probability flux vector field, culminating in the presentation of the multidimensional rate constant in equation \eqref{Langer_rate_constant}. Lastly, Section \ref{BS_simplification} discusses some more recent development by Berezhkovskii and Szabo \citep{berezhkovskii_one-dimensional_2004} that allows one to project Langer's result back to one dimension. Together, the results presented in this note are sometimes referred to as the KLBS theory.

Given the venerable age of Langer's theory and the existence of several review articles \citep{hanggiReactionrateTheoryFifty1990, zhouRateTheoriesBiologists2010, petersReactionCoordinatesMechanistic2016} and textbook \citep{petersReactionRateTheory2017} partly devoted to this subject, one might wonder why this note is necessary. This introduction certainly draws heavily on the aforementioned works, but differs in two regards. First, almost all current accounts of Langer's theory are quite concise. While this feature is excellent for experts who are in need of a quick refresher, it poses significant challenges for beginners, because the derivation of the theory can be quite tedious and involve some mathematical tricks that are not standard knowledge for a typical reader. One goal of this note, therefore, is to derive Langer's theory while erring on the side of presenting too many, rather than too few, algebraic details.

Second, an interesting feature of Langer's theory is that the expression for the rate constant critically depends on several basic geometric features of the committor/splitting probability near the saddle point. This connection was not explicitly stated in Langer's original paper, and is usually not fully developed in recent accounts of the theory. This is, again, unfortunate, since this connection provides a simple setting for gaining an intuitive understanding of the behavior of the committor near an idealized transition state. Given the prominent role the committor plays in modern rate theories, such as the transition path theory \citep{eTransitionPathTheoryPathFinding2010}, such intuitions can be valuable in understanding the more recent development. Therefore, as described above, a second goal of this note is to make explicit the connection of Langer's theory to the committor function and provide a detailed analysis thereof.

The primary intended audience for this note are students and researchers in chemistry and biophysics interested in condensed-phase simulations of macromolecules. The reader should be familiar with linear algebra and have some exposure to classical statistical mechanics and some rate theory (e.g., Arrhenius equation); knowledge of stochastic processes is not strictly necessary although results from the theory will be invoked in some derivations.

\section{A brief review of stochastic process}
\label{review_section}

In this section we briefly review some concepts from the theory of stochastic process relevant for our discussion. Readers who have never studied this subject should either consult standard references on stochastic processes and statistical mechanics, or simply take the results presented below as given and fill in their missing background knowledge later. The presentation in this section partially follows that in Chapter 15 of \citep{tuckermanStatisticalMechanicsTheory2010}.

\subsection{Overdamped Langevin equation}
\label{review_langevin}

Consider a (closed) system that is in contact with a large heat bath. In a typical (classical) simulation, the system consists of the macromolecule(s) of interest as well as some water molecules and counterions. The time evolution of the full system is determined by the Hamiltonian equations, and the equilibrium phase space distribution function is microcanonical. To avoid explicit representation of the bath, we ``abstract away'' the bath degrees of freedom, and the remaining system dynamics can be described via the generalized Langevin equation:
\begin{align}
 r_i'(t) &= \frac{p_i(t)}{m_i} \\
 p_i'(t) &= -\frac{\partial W(r)}{\partial r_i} - \int_0^t \sum_j \sqrt{m_im_j}\gamma_{ij}(t-\tau)r_j'(\tau)\,d\tau + \xi_i(t) \label{GLE_v'}
\end{align}
Here, $r_i$, $p_i$, and $m_i$ are the position, momentum, and (renormalized) mass of the $i$th degree of freedom in the remaining system, and $W(r)$ is the potential of mean force obtained from the full potential energy function by averaging over the bath degrees of freedom. The generalized Langevin equation can be derived via either the harmonic bath model or, more rigorously, via the Mori–Zwanzig theory.

The generalized Langevin equations is reminiscent of Newton's equation of motion, but the price we pay for doing away with the bath degrees of freedom is the difficulty of dealing with a set of stochastic integro-differential equations. Specifically, compared to Newton's equations, there are two additional terms in \eqref{GLE_v'} that provide a coarse-grained model of the effect of the bath on the system. Here, we have ``decomposed'' the effect of the bath into two parts. First, the fluctuation term $\xi_i(t)$ represents a random force (i.e., noise) acting on the system. Although the motion of the bath is fully deterministic, by ignoring the molecular details we model $\xi_i(t)$ as a random, or stochastic, process. For a system solvated by a dense bath, such as liquid water, that affects the system dynamics through soft collisions (i.e., weak noise), a common model for $\xi_i(t)$ is a Gaussian random process with zero mean.

Second, the dissipation term $\int_0^t \gamma_{ij}(t-\tau)r_j'(\tau)\,d\tau$ is a convolution integral that acts as a friction force slowing down the system. This integral is called a memory integral; the term $\gamma_{ij}(t)$ is known as the memory kernel, or dynamic friction kernel, and it encodes the ``memory'' of the motion of the system by the bath. Physically, the memory integral represents the fact that the bath requires a finite amount of time to respond to fluctuations in the system and this lag affects the motion of the system. At equilibrium, the fluctuation term and the dissipation term are related by the second fluctuation-dissipation theorem,
\begin{equation}
 \mathbb E[\xi_i(t)\xi_j(t')] = k_BT\sqrt{m_im_j}\gamma_{ij}(|t-t'|) \label{GLE_colored_noise}
\end{equation}

Now, we make some further simplifying assumptions to make \eqref{GLE_v'} more analytically tractable. First, we assume that the bath responds instantaneously to the motion of the system; i.e., we assume that the memory integral decays instantaneously and the bath has no memory of the system history. This is a good model when the (renormalized) mass of the system is much larger than that of the bath. In this case, the memory kernel becomes $\gamma_{ij}(t)=2\gamma_{ij}\delta(t)$, where we have defined
\begin{equation}
 \gamma_{ij}=\int_0^\infty\gamma_{ij}(t)\,dt \label{static_friction_kernel}
\end{equation}
as the static friction kernel, or simply the friction coefficient, and $\delta(t)$ is a Dirac delta function. The fluctuation-dissipation theorem now reads
\begin{equation}
 \mathbb E[\xi_i(t)\xi_j(t')] = 2k_BT \sqrt{m_im_j}\gamma_{ij}\delta(|t-t'|) \label{GLE_white_noise}
\end{equation}
Stochastic processes with an autocorrelation function of this form are called white noises, and the terminology reflects the fact that the power spectral density of the process is a constant over all frequencies. With these assumptions, the generalized Langevin equation becomes
\begin{align}
 r_i'(t) &= \frac{p_i(t)}{m_i} \\
 p_i'(t) &= -\frac{\partial W(r)}{\partial r_i} - \sum_j\sqrt{m_im_j}\gamma_{ij}r_j'(t) + \xi_i(t) \label{LE_v'}
\end{align}
which is known as the Langevin equation. Compared to the generalized Langevin equation, the Langevin equation describes stochastic processes that are Markovian (i.e., memoryless).

Second, for dense solvent, such as water, the high friction and frequent collisions with the system leads to, on a short timescale (i.e., $t<\gamma^{-1}$), rapid fluctuations in the acceleration $p_i'(t)$. However, on a longer timescale the change in the time-averaged velocity will be small as the effect of collisions cancel out each other. This allows us to set the acceleration in \eqref{LE_v'} to zero, which leads to
\begin{equation}
 \sum_j\sqrt{m_im_j}\gamma_{ij}r_j'(t) = -\frac{\partial W(r)}{\partial r_i} + \xi_i(t) \label{overdamped_LE}
\end{equation}
The motion described by \eqref{overdamped_LE} goes by many names, such as diffusion, Brownian motion, or overdamped Langevin dynamics; the equation itself is known as the overdamped Langevin equation. In some applications, the cross-correlation terms in the memory kernel are ignored, in which case \eqref{overdamped_LE} takes on a simpler form
\begin{equation}
 m_i\gamma_{ii}r_i'(t) = -\frac{\partial W(r)}{\partial r_i} + \xi_i(t)
\end{equation}

\subsection{Smoluchowski equation}

For the purpose of describing the kinetics of activated rate processes at equilibrium, working directly with \eqref{overdamped_LE} is inconvenient, since trajectories consistent with \eqref{overdamped_LE} are individual realizations of the system dynamics, while we are more interested in the statistics of an ensemble of such realizations. In other words, we are more interested in $p(r, t)$, the probability density of the system (strictly speaking, $p(r, t)$ is a conditional probability density function more appropriated denoted as $p(r, t | r_0, t_0)$). The time evolution of $p(r, t)$ for processes governed by \eqref{overdamped_LE} is given by
\begin{equation}
\frac{\partial p(r, t)}{\partial t}=-\nabla\cdot J(r, t) \quad\text{with}\quad J(r, t) = -D\pi(r)\nabla\frac{p(r, t)}{\pi(r)} \label{Langer_FP}
\end{equation}
where $\pi(r)$ is the Boltzmann distribution in the configuration space and the stationary solution to \eqref{Langer_FP}, $J(r, t)$ is the (probability) flux, and $D$ is the (position-independent) diffusion matrix, which we assume to be symmetric positive definite. Elements of the diffusion matrix are related to the static friction kernel defined in \eqref{static_friction_kernel} via the relation
\begin{equation}
 D_{ij}=\frac{k_BT}{\sqrt{m_im_j}\gamma_{ij}}
\end{equation}
Equations such as \eqref{Langer_FP} are known as the Smoluchowski equation, a special case of a class of partial differential equations known as the Fokker-Planck equations, or Kolmogorov’s forward equations. It is also common to refer to the first part of \eqref{Langer_FP} as the continuity equation for the probability density, and the second part of \eqref{Langer_FP} as a ``constitutive'' equation.

For the following discussion it is also convenient to rewrite \eqref{Langer_FP} in two forms. First, we can rewrite \eqref{Langer_FP} component-wise:
\begin{equation}
 \partial_tp(r, t)=\sum_{ij}\partial_iD_{ij}\left[\beta\partial_j W(r) + \partial_j\right]p(r, t) \label{Langer_FP_elementwise}
\end{equation}
where we have used a short-hand notation $\partial_t=\partial/\partial t$ and $\partial_i=\partial/\partial r_i$. For readers not familiar with the component-wise notation, they should convince themselves that, e.g., the $i$th component of a matrix-vector product $Ax$ can be expressed as $\sum_j A_{ij}x_j$.

Second, we can rewrite \eqref{Langer_FP} as an equation involving the Fokker-Planck operator $L^\dagger$,
\begin{equation}
 \frac{\partial p(r, t)}{\partial t}=L^\dagger p(r, t) \quad\text{where}\quad L^\dagger=\nabla\cdot D\pi(r)\nabla\pi^{-1}(r) \label{Langer_FP_operator}
\end{equation}
Here, the symbol $\dagger$ indicates that the Fokker-Planck operator $L^\dagger$ is the adjoint of another operator $L$ known as the generator,
\begin{equation}
 L=\pi^{-1}(r)\nabla\cdot D\pi(r)\nabla \label{Langer_generator}
\end{equation}
The concept of an adjoint operator is a generalization of the Hermitian transpose of a matrix. As \eqref{Langer_FP_operator} shows, the Fokker-Planck operator dictates the time evolution of the probability density $p(r, t)$. There is a similar interpretation for the generator: the generator dictates the time evolution of (conditional) ensemble averages, or observables, of the form $u(r, t)=\mathbb E[f(r(t))|r(0)=r_0]$, for some suitable scalar function $f$ defined on the configuration space; i.e.,
\begin{equation}
 \frac{\partial u(r, t)}{\partial t}=Lu(r, t)
\end{equation}

\section{Kramers' theory at the high friction limit}
\label{Kramers_theory}

In this section we consider Kramers' theory in the context of a one-dimensional Brownian particle whose motion is described by the one-dimensional version of the overdamped Langevin equation \eqref{overdamped_LE},
\begin{equation}
 m\gamma r'(t)=-\frac{dW}{dr}+\xi(t) \quad\text{where}\quad \mathbb E[\xi(t)]=0, \mathbb E[\xi(t)\xi(t')] = 2k_BT m\gamma\delta(|t-t'|) \label{1D_overdamp_langevin}
\end{equation}
From \eqref{Langer_FP}, it follows that the probability density $p(r, t)$ satisfies the following Fokker-Planck equation
\begin{equation}
 \frac{\partial p(r, t)}{\partial t}=-\frac{\partial J(r, t)}{\partial r} \quad\text{with}\quad J(r, t)=-D\left[\beta\frac{dW}{dr}+\frac{\partial}{\partial r}\right]p(r, t) \label{Kramers_FP}
\end{equation}
where $D=k_BT/m\gamma$ is the position-independent diffusion constant. The stationary solution to \eqref{Kramers_FP} is the Boltzmann distribution $\pi(r) = e^{-\beta W(r)}/Z$, where $Z$ is a configurational partition function.

\begin{figure}[b!]
    \centering
    \includegraphics[width=2.5in]{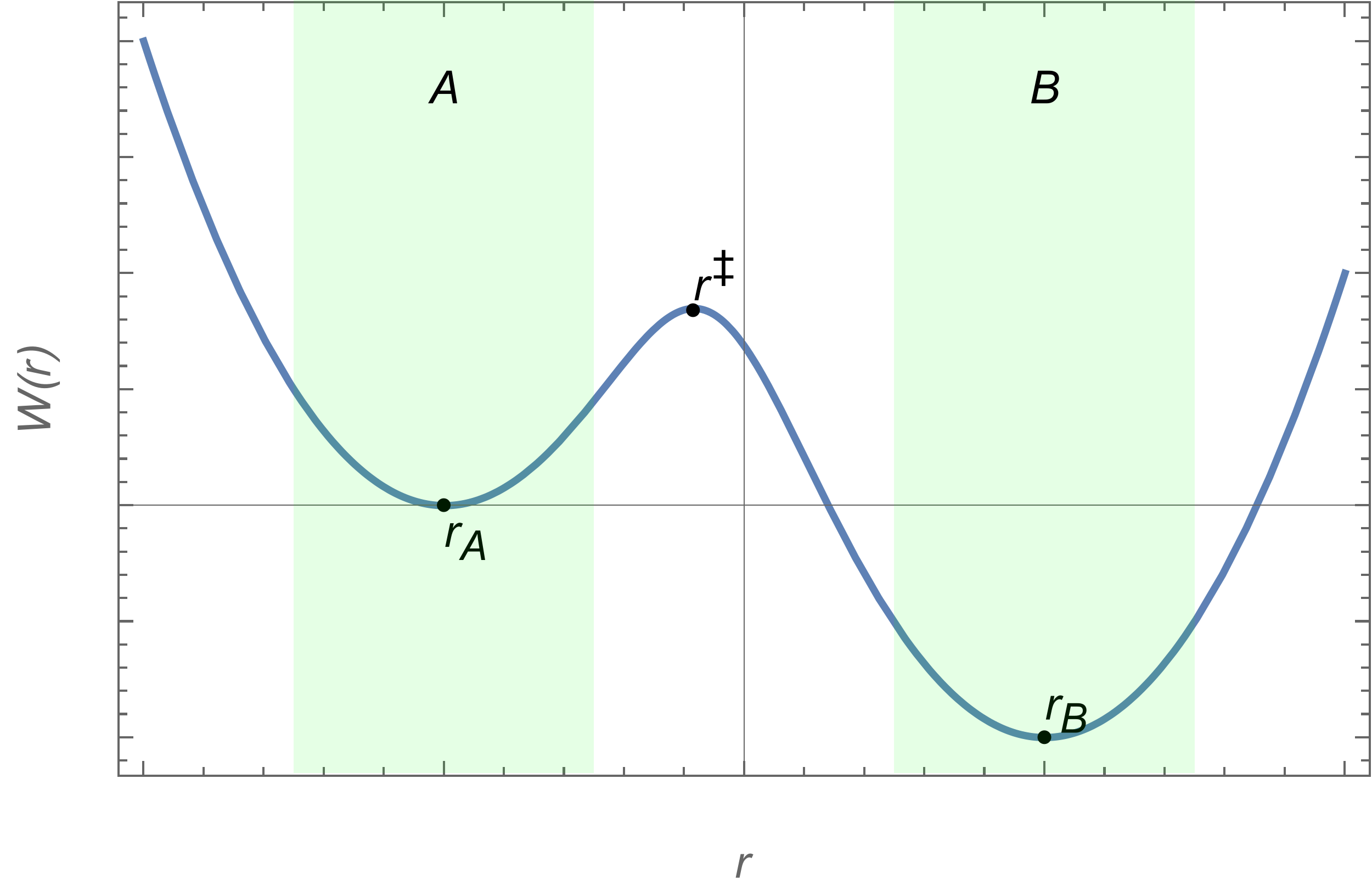}
    \caption{An example one-dimensional double-well potential of mean force. The potential is given by $W(r)=-\ln\left(e^{-2(x+1)^2}+e^{-2(x-1)^2+1}\right)$.}
    \label{fig:1d_pmf}
\end{figure}

Here, we take $W(r)$ to be an (asymmetric) double-well potential, where the minimum of the reactant well $A$ is located at $r_A$, the minimum of the product well $B$ is located at $r_B$, and the transition state $r^\ddagger$ is identified as the position of the peak of $W(r)$ between the two wells; without loss of generality, we assume that $r_A<r^\ddagger<r_B$. An example of such a potential is shown in Figure \ref{fig:1d_pmf}. Further, we assume that the height of the barrier $W(r^\ddagger)$ is much larger than $k_BT$, so that there is a separation of timescales between barrier crossing and within-well equilibration. A related assumption here is that barrier crossing is much slower than the correlation time of the dynamic friction kernel so that the static friction kernel approximation can be justified. Together, these assumptions amount to the situation where a single slow degree of freedom in the system $r$ is sufficient for describing the reaction, while the rest of the degrees of freedom relax much faster than the timescale of barrier crossing along $r$; we should thus interpret ``bath'' as also including other degrees of freedom of the macromolecule that are not explicitly treated, and interpret $r$ not necessarily as a position variable but more generally as a reaction coordinate (i.e., some function of the position variables $r_i$).

We mention here in passing that Kramers also derived expressions for the rate constant in the weak and moderate-to-high friction regimes. We will not discuss these results since they are not as relevant for condensed-phase simulations.

\subsection{Kramers' theory via the flux-over-population method}

Kramers derived an expression for the rate constant of the $A\to B$ reaction, $k_{AB}$, using what is now known as the flux-over-population method. Consider a hypothetical procedure in which an ensemble of $n$ particles are prepared in the reactant well $A$ (usually we set $n=1$), where they rapidly reach thermal equilibrium on a timescale much faster than that of escaping to the product well $B$. Whenever a particle escapes from the reactant well and reaches the product well, it is immediately removed and a new one is added to the reactant well, such that the reactant population is always maintained at $n$. As we show in the analysis below, the exact positions at which a particle is considered escaped and at which a new one is inserted are not very important. As the system reaches a non-equilibrium steady-state, there is a non-zero probability current $J$, or flux, into the product well, which is the reaction rate (i.e., the number of transitions from $A$ to $B$ per unit time). We normalize this reaction rate by the reactant population, which gives the rate constant,
\begin{equation}
 k_{AB}=J/n \label{flux_over_pop}
\end{equation}
hence the name ``flux-over-population''.

Before flushing out the algebraic details, let us pause for a moment and consider why the flux-over-population procedure is needed. In other words, why does the calculation of an equilibrium rate constant invoke the flux of a seemingly contrived non-equilibrium process? A simple answer is that the (net) flux between any two states (micro- or macro-) in equilibrium is zero, due to detailed balance, and thus gives no information about kinetics. To see why the flux-over-population procedure circumvent this problem, let us consider the behavior of particles at some point near the transition state and see how they contribute to the flux across that point at equilibrium. At any given moment, such a particle can be categorized into one of four groups, depending on its past and future behavior:
\begin{enumerate}
 \item The particle came from $A$ and will move to $B$ before going back to $A$.
 \item The particle came from $A$ and will go back to $A$ before moving to $B$.
 \item The particle came from $B$ and will move to $A$ before going back to $B$.
 \item The particle came from $B$ and will go back to $B$ before moving to $A$.
\end{enumerate}
Behavior described in group 2 and 4 are known as barrier recrossing and does not contribute to the flux, since each pair of crossing-recrossing cancel out each other (this is to be contrasted with the procedure in transition-state theory, where all (re)crossing events are counted towards the total flux across some dividing surface \citep{vanden-eijndenTransitionStateTheory2005}). Because the net equilibrium flux is zero, this implies that the flux contributed by group 1 and 3 cancels as well. Note that, for ergodic dynamics, this categorization is exhaustive and it is not possible for a particle to cross the barrier once and stay in one of the two wells forever. By removing particles reaching the product well, the flux-over-population procedure removes the flux contribution from group 3 (and group 4), while the only remaining nonzero flux contribution from group 1 stays close to its equilibrium value at the steady state due to the separation of timescales. As such, the nonequilibrium steady-state flux $J$ in this procedure gives the true $A$ to $B$ reaction rate.

At steady state, the population distribution $p_\text{ss}(r)$ is time-independent, and thus by \eqref{Kramers_FP} the flux $J$ is now both time- and position-independent. The two quantities are related via
\begin{equation}
 J = -D\left[\beta\frac{dW}{dr}+\frac{\partial}{\partial r}\right]p_\text{ss}(r)= -De^{-\beta W(r)}\frac{d}{dr}\left[e^{\beta W(r)}p_\text{ss}(r)\right]
\end{equation}
Now, divide both sides by $De^{-\beta W(r)}$ and integrate both sides  over the interval $[r_A, r_B]$, which gives
\begin{equation}
J\int_{r_A}^{r_B}D^{-1}e^{\beta W(r)}\,dr = - e^{\beta W(r)}p_\text{ss}(r)\big|_{r=r_A}^{r_B} \label{Kramers_flux}
\end{equation}
We now consider the behavior of $p_\text{ss}(r)$ near the two boundaries: $r_A$ and $r_B$. At $r=r_B$, we require $p_\text{ss}(r_B)=0$ to satisfy the absorbing boundary condition caused by particle removal. At positions in the reactant well away from the absorbing boundary condition at $r_B$, we assume that the system is close to equilibrium, i.e., $p_\text{ss}(r)\approx\pi(r)$, because the particle insertion procedure maintains the reactant population $n$ in the reactant well, and inserted particles thermalize much faster than the reaction timescale. This observation has two consequences. First, it implies that $p_\text{ss}(r_A)=\pi(r_A)$. Second, it implies that the population in the reactant well is
\begin{equation}
 n= \int_{-\infty}^{r^\ddagger}p_\text{ss}(r)\,dr\approx \int_{-\infty}^{r^\ddagger}\pi(r)\,dr
\end{equation}
Now, apply the boundary conditions at $p_\text{ss}(r_A)$ and $p_\text{ss}(r_B)$ to \eqref{Kramers_flux}, then apply the flux-over-population formula \eqref{flux_over_pop}, and we arrive at
\begin{align}
 k_{AB} &= \frac{e^{\beta W(r_A)}p_\text{ss}(r_A) - e^{\beta W(r_B)}p_\text{ss}(r_B)}{\int_{r_A}^{r_B}D^{-1}e^{\beta W(r)}\,dr \int_{-\infty}^{r^\ddagger}p_\text{ss}(r)\,dr} \\
 &= \frac{e^{\beta W(r_A)}\pi(r_A)}{\int_{r_A}^{r_B}D^{-1}e^{\beta W(r)}\,dr \int_{-\infty}^{r^\ddagger}\pi(r)\,dr} \\
 &= \frac{1}{\int_{r_A}^{r_B}D^{-1}e^{\beta W(r)}\,dr \int_{-\infty}^{r^\ddagger}e^{-\beta W(r)}\,dr} \label{Kramers_rate_1}
\end{align}

If we assume that $W(r)$ is harmonic around $r^\ddagger$ and $r_A$, we write, for $W(r)$ near $r^\ddagger$ and $r_A$,
\begin{equation}
 W(r)\approx\Delta W^\ddagger -\frac 1 2 \kappa^\ddagger\left(r-r^\ddagger\right)^2 \quad\text{and}\quad W(r)\approx \frac 1 2 \kappa_A\left(r-r_A\right)^2 \label{Kramers_harmonic}
\end{equation}
where we have defined $\Delta W^\ddagger= W(r^\ddagger)-W(r_A)$ and set $W(r_A)=0$ (resetting the zero of the potential of mean force has the effect of changing the partition function, which has no effect on our derivation). Here, $\kappa_A$ and $\kappa^\ddagger$ are the force constants of the harmonic potentials; equivalently, $\kappa_A$ and $\kappa^\ddagger$ can also be interpreted as the curvatures at $r_A$ and $r^\ddagger$, respectively. Applying \eqref{Kramers_harmonic} to \eqref{Kramers_rate_1} gives the following approximate results
\begin{align}
 \int_{r_A}^{r_B}D^{-1}e^{\beta W(r)}\,dr &\approx \int_{-\infty}^\infty D^{-1}e^{\beta\left(\Delta W^\ddagger - \kappa^\ddagger r^2/2\right)}\,dr =  D^{-1}\sqrt{\frac{2\pi}{\beta\kappa^\ddagger}} e^{\beta\Delta W^\ddagger} \label{Kramers_int_saddle}\\
 \int_{-\infty}^{r^\ddagger}e^{-\beta W(r)}\,dr &\approx \int_{-\infty}^\infty e^{-\beta\kappa_Ar^2/2}\,dr = \sqrt{\frac{2\pi}{\beta\kappa_A}} \label{Kramers_int_reactant}
\end{align}
These results provide a good approximation whenever the region of validity for the harmonic approximation is large enough that the added probability mass by taking the integration limit to infinity is negligible.

Substituting \eqref{Kramers_int_saddle} and \eqref{Kramers_int_reactant} back to \eqref{Kramers_rate_1} gives
\begin{equation}
 k_{AB}=\frac{\beta D}{2\pi}\sqrt{\kappa_A\kappa^\ddagger} e^{-\beta\Delta W^\ddagger} \label{Kramers_rate_2}
\end{equation}
This is the rate constant predicted by Kramers' theory at the high friction limit. Since the dynamics at high friction is diffusive, it is also known as the spatial-diffusion-limited rate. The rate constant for the reverse reaction can be obtained analogously by replacing the reactant well $A$ with the product well $B$ in the preceding derivation. If the diffusion constant is position-dependent, the derivation up to (and include) \eqref{Kramers_rate_1} is still valid. The diffusion constant can be ``folded'' into the potential of mean force $W(r)$ as
\begin{equation}
 e^{\beta W(r)}/D(r)=e^{\beta W(r) -\ln D(r)}=e^{\beta\left[W(r)-k_BT\ln D(r)\right]}
\end{equation}
This defines a new potential surface $W(r)-k_BT\ln D(r)$, which may reach its maximum at a position other than $r^\ddagger$.

In the literature it is also common to parameterize the harmonic potential near $r^\ddagger$ and $r_A$ as
\begin{equation}
 W(r)\approx\Delta W^\ddagger -\frac 1 2 m {\omega^\ddagger}^2\left(r-r^\ddagger\right)^2 \quad\text{and}\quad W(r)\approx \frac 1 2 m \omega_A^2\left(r-r_A\right)^2 \label{Kramers_harmonic_alt}
\end{equation}
Here, $\omega_A$ (and analogously, $\omega^\ddagger$) is the angular frequency of a harmonic oscillator of mass $m$ on the potential surface $W(r)=m\omega_A^2r^2/2$; $\omega_A$ is related to the force constant/curvature $\kappa_A$ by $\kappa_A=m{\omega_A}^2$. Together with the definition of the diffusion constant $D=1/\beta m\gamma$, \eqref{Kramers_rate_2} can be written alternatively as
\begin{equation}
 k_{AB}=\frac{\omega_A\omega^\ddagger}{2\pi\gamma}e^{-\beta\Delta W^\ddagger} \label{Kramers_rate_alt}
\end{equation}

\subsection{Kramers' theory via MFPT}

There is an alternative approach to deriving \eqref{Kramers_rate_2} using the mean first-passage time (MFPT). Let $\Omega$ be the reaction coordinate space of the system. For a subset $B\subseteq\Omega$, the MFPT $m_B(r)$ is the average time for the system initiated at $r$ to reach $B$. It can be shown, using potential theory, that the MFPT satisfies the following (Dirichlet) boundary value problem,
\begin{equation}
 \begin{cases}
  Lm_B(r)=-1, & r\in\Omega\backslash B \\
  m_B(r)=0, & r\in B
 \end{cases}
 \label{MFPT_Dirichlet}
\end{equation}
where the notation $\Omega\backslash B$ denotes the complement of $B$ in $\Omega$. For one-dimensional Brownian dynamics, the generator $L$ defined in \eqref{Langer_generator} takes on the form
\begin{equation}
 L= \frac{d}{dr}D\left[\frac{d}{dr} - \beta W(r)\right]= D\frac{d^2}{dr^2} - D\beta\frac{dW}{dr}\frac{d}{dr}
\end{equation}
Unfortunately, a full derivation of \eqref{MFPT_Dirichlet} from scratch would take too long; interested readers should consult standard references on Markov processes for more details. Nevertheless, an ``intuitive'' justification of \eqref{MFPT_Dirichlet} goes as follows: consider a hypothetical procedure where an ensemble of system trajectories are prepared, all with the same initial condition at position $r(0)=r_0$ outside $B$; we can calculate $m_B(r_0)$ by taking the average first hitting time to the set $B$ among the trajectories. For each given trajectory $r(t)$, moving forward in time by a small amount $\delta t$ reduces the first hitting time at $r(t+\delta t)$ by the same amount of $\delta t$; as such, the time derivative of $m_B(r_0)$ should be $-1$. Since the generator acting on an observable $m_B(r_0)$ gives its time derivative, it follows that $Lm_B(r_0)=-1$. On the other hand, if $r_0$ is already in $B$, then the first hitting time to $B$ is zero by construction, hence the boundary condition in \eqref{MFPT_Dirichlet}.

\begin{figure}[t!]
    \centering
    \includegraphics[width=2.2in]{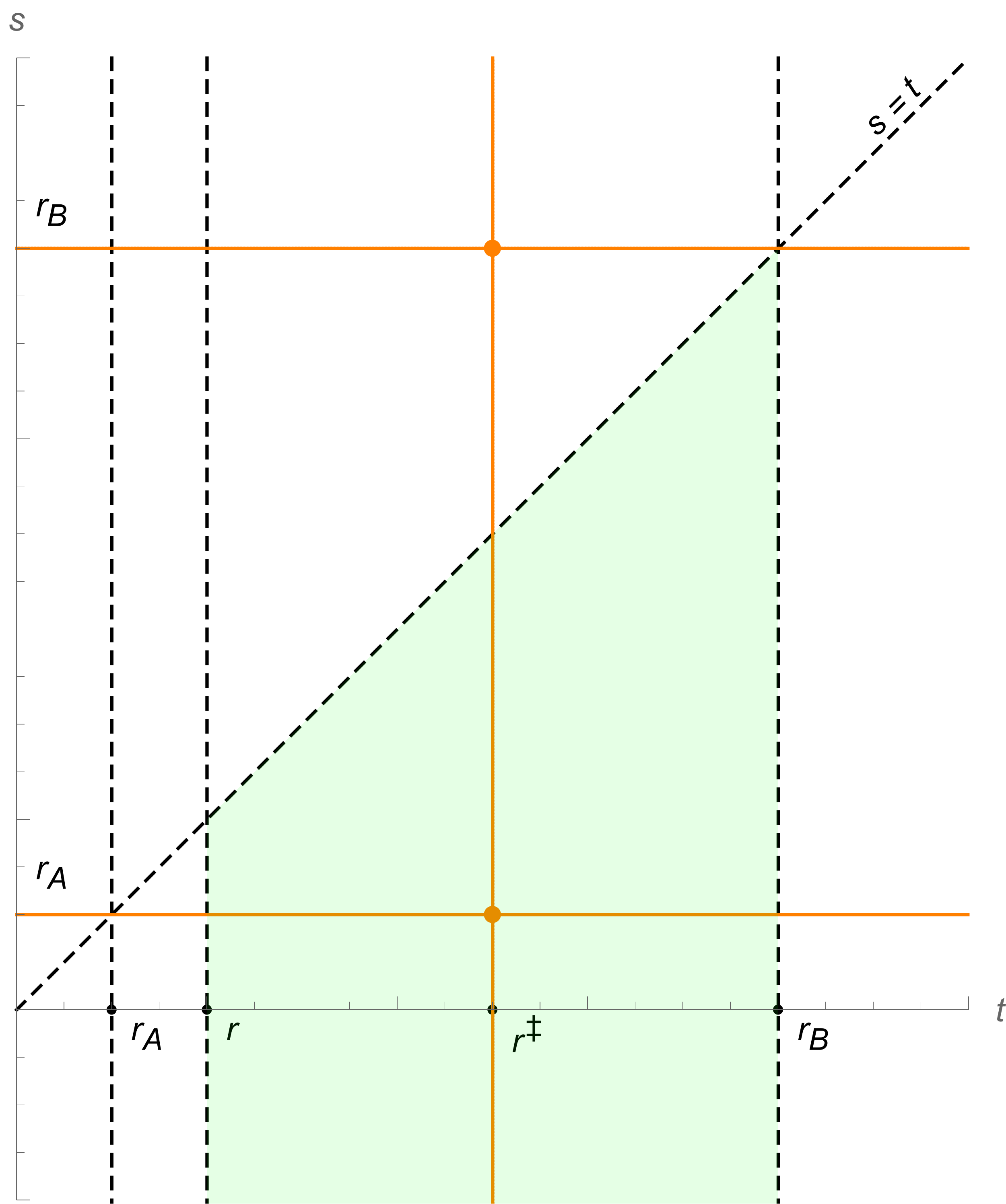}
    \caption{The region of integration for the double integral in \eqref{Kramers_MFPT}. The region of integration is shown in green. The maxima of $e^{\beta W(t)}e^{-\beta W(s)}$ are shown as intersections of the orange lines. Only one of the two maxima, $(s=r_A, t=r^\ddagger)$, is located inside the integration limit.}
    \label{fig:mfpt_int_region}
\end{figure}

To determine the rate constant for the transition from the reactant well to the product well, let us consider the MFPT to the product well from some point $r$ near the reactant well (as we will see, the particular choice of $r$ is not very important). More specifically, let $\Omega=\mathbb R$ and consider the MFPT for the set $B=[r_B,\infty)$. The differential equation $Lm_B(r)=-1$ can be solved using an integrating factor,
\begin{align}
 D\frac{d^2m_B}{dr^2} - D\beta\frac{dW}{dr}\frac{dm_B}{dr} &= -1\\
 \frac{d}{dr}\left[e^{-\beta W(r)}\frac{dm_B}{dr}\right] &= -D^{-1}e^{-\beta W(r)} \\
 \int_{-\infty}^{t}\frac{d}{ds}\left[e^{-\beta W(s)}\frac{dm_B}{ds}\right]\,ds &= -D^{-1}\int_{-\infty}^t e^{-\beta W(s)}\,ds \\
 \int_r^{r_B}\frac{dm_B}{dt}\,dt &= -\int_r^{r_B} D^{-1}e^{\beta W(t)}\int_{-\infty}^t e^{-\beta W(s)}\,dsdt \\
 m_B(r) &= \int_r^{r_B} D^{-1}e^{\beta W(t)}\int_{-\infty}^t e^{-\beta W(s)}\,dsdt \label{Kramers_MFPT}
\end{align}
It is easy to check that, under the harmonic approximations of \eqref{Kramers_harmonic}, the function $e^{\beta W(t)}e^{-\beta W(s)}$ has a unique global maximum at $(s=r_A, t=r^\ddagger)$ within the domain of integration in \eqref{Kramers_MFPT} (see Figure \ref{fig:mfpt_int_region}). Since $e^{\beta W(t)}e^{-\beta W(s)}$ decays exponentially away from the global minimum, we can simply extend the domain of integration to $\mathbb R^2$, in which case the double integral in \eqref{Kramers_MFPT} simplifies to the integrals considered in \eqref{Kramers_int_saddle} and \eqref{Kramers_int_reactant}. Note that the validity of this approximation implies that $m_B(r)$ is a constant near and to the left of the reactant well, and that the particular choice for the boundaries of $B$ is not important as long as its lower bound is sufficiently to the right of $r^\ddagger$. 

Taken together, we have shown that the rate constant \eqref{Kramers_rate_2} derived from the flux-over-population method is equivalent to the inverse of the mean first-passage time; in fact, this equivalence is exact for a much broader range of stochastic processes than those considered in this note \citep{reimannUniversalEquivalenceMean1999}. Unfortunately, despite its conceptual simplicity, the MFPT cannot in general be calculated analytically for problems in more than one dimension and thus we will not consider this method in the following sections; however, see Section VII. D. of \citep{hanggiReactionrateTheoryFifty1990} for an approximate approach to solving for the MFPT in higher dimensions using asymptotic methods.

\section{Langer's extension to higher dimensions}
\label{Langer_theory}

\begin{figure}[t!]
    \centering
    \includegraphics[width=7.1in]{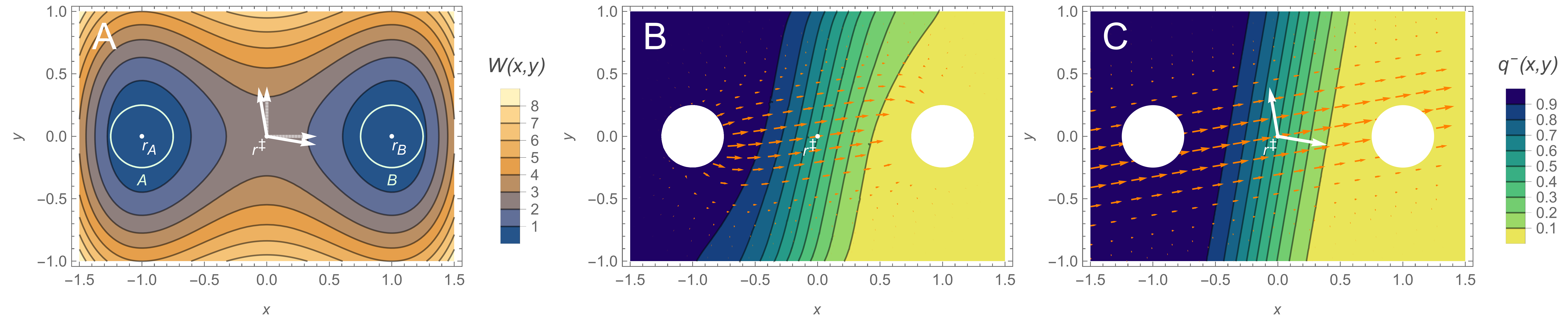}
    \caption{An example two-dimensional double-well potential. A) Contour plot of the double-well potential; the potential is given by $W(x,y)=\frac{5}{2}(x^2-1)^2+5y^2$ (see also \citep{metznerIllustrationTransitionPath2006}). The reactant well $A$ and the product well $B$ are defined, somewhat arbitrarily, as disks of radius $1/4$ centered at the global minima $r_A$ and $r_B$. The white dashed vectors show the eigenvectors of $H^\ddagger$, while the white solid vectors show the eigenvectors of $H^\ddagger D$, with $D=\{\{4, 1\}, \{1, 2\}\}$. B) The exact backward committor $q^-(x, y)$ and steady-state flux $J(x, y)$. The committor is computed by solving \eqref{Langer_committor_eq} numerically using the \texttt{NDSolveValue} function with the appropriate \texttt{DirichletCondition} in Mathematica 12.1 and is shown as a contour plot (with 0.1 contour increment). The steady-state flux is calculated using \eqref{Langer_ss_flux_def} and represented as the orange vector field; the lengths of the streamline segments are proportional to the vector field magnitude. C) The approximate backward committor and steady-state flux from Langer's theory. The committor is given by \eqref{Langer_isocommittor_solution}, and the steady-state flux is given by \eqref{Langer_ss_flux} with $\pi(s)$ replaced with its harmonic approximation at the saddle point. For all calculations $\beta$ is set to one.}
    \label{fig:2d_example}
\end{figure}

Now let us consider the extension of Kramers' theory to an $N$-dimensional potential (of mean force) with the reactant and product wells separated by a saddle point; recall that the saddle point is a point on the potential energy surface $W(r)$ where the gradient is zero and the Hessian has exactly one unstable mode. An example two-dimensional double-well potential is shown in Figure \ref{fig:2d_example}A. Conceptually, the flux-over-population method employed in the one-dimensional case carries over with few changes. However, as we will see, the higher-dimensional setting of \eqref{Langer_FP} does pose some algebraic challenges. The derivation presented here loosely follows the presentation in \citep{langerStatisticalTheoryDecay1969, hanggiReactionrateTheoryFifty1990, zhouRateTheoriesBiologists2010}.

Similar to our treatment in one dimension, we assume a harmonic approximation at $r_A$, the bottom of the reactant well $A$,
\begin{equation}
 W(r)\approx \frac 1 2 (r-r_A)^TH_A(r-r_A) \label{Langer_harmonic_reactant}
\end{equation}
where $H_A$ is the Hessian matrix of second-order derivatives evaluated at the minimum $r_A$, which we assume to be symmetric positive definite; note that the Hessian is also known as the force constant matrix (to see why, note that for a harmonic potential $\kappa r^2/2$ with force constant $\kappa$, the second-order derivative at the center $r=0$ is simply $\kappa$). In terms of notations, in this note we will exclusively use notations of the form $x^Ty$ to denote inner products in place of other common notations such as $x\cdot y$ or $\left<x, y\right>$; readers more familiar with the other notations should convince themselves that, e.g., the quadratic form in \eqref{Langer_harmonic_reactant} can be expressed in the angle bracket notation as $\left<r-r_A, H_A(r-r_A)\right>/2$.

In addition, we assume that there is a saddle point $r^\ddagger$ in between the reactant and product wells, which is the point with the minimum energy along the barrier ridge. A harmonic approximation at the saddle point gives
\begin{equation}
 W(r) \approx \Delta W^\ddagger + \frac 1 2 (r-r^\ddagger)^TH^\ddagger(r-r^\ddagger) \label{Langer_harmonic_saddle}
\end{equation}
Here, we assume that $H^\ddagger$ has exactly one negative eigenvalue associated with the unstable mode at $r^\ddagger$, while the rest of the eigenvalues are assumed to be strictly positive. In general, it is possible for $H^\ddagger$ to possess one or more eigenvalues of zero, which correspond to some underlying symmetries of the system. For example, the Hessian matrix of a system of two one-dimensional particles with positions $x_1$ and $x_2$ whose distance is subject to the harmonic restraint $W(x_1, x_2)=\kappa(r_0 - (x_1-x_2))^2/2$ will have a zero eigenvalue corresponding to the translational symmetry of the system. Such symmetries can oftentimes be eliminated (e.g., here by working with the reaction coordinate $r=x_1-x_2$) and we shall not consider such cases for the following analysis. Lastly, we record here for future reference,
\begin{equation}
 \partial_i W(r)= \sum_j H^\ddagger_{ij}\left(r_j-r^\ddagger_j\right) \label{Langer_quadratic_form_derivative}
\end{equation}
This result comes from the fact that $H^\ddagger$ is symmetric, and, for any quadratic form $x^TAx$, $\nabla_x x^TAx=(A+A^T)x$.

\subsection{A relation between the steady-state probability and the backward committor}

Instead of seeking a steady-state solution $p_\text{ss}(r)$ to \eqref{Langer_FP} directly, we take an indirect approach that relates $p_\text{ss}(r)$ to a function called the backward committor $q^-(r)$, which is the probability that, going back in time, a trajectory starting at $r$ will reach the reactant well before the product well. As we will see, $p_\text{ss}(r)$, a nonequilibrium quantity, is related to the behavior of $q^-(r)$ at equilibrium by a surprisingly simple equation. Since Brownian dynamics under detailed balance is time-reversible (in the sense that a time-reversed process is statistically indistinguishable from the original one), it is more convenient in this context to think of $q^-(r)$ as representing the probability that, going forward in time, a trajectory starting at $r$ will reach the reactant well before the product well. The choice of the superscript notation is intended to distinguish $q^-$ from $q^+$, the forward committor, which is the probability that, going forward in time, a trajectory starting at $r$ will reach the product well before the reactant well; for time-reversible processes at equilibrium the two quantities are related by $q^+(r)+q^-(r)=1$. The committor is also known as splitting probability, or, in the protein folding literature, $p_\text{fold}$. One should be careful not to confuse the definition of time reversibility in the theory of stochastic process with the concept of reversibility in Hamiltonian dynamics (i.e., flipping both time and momentum leaves the process invariant), the concept of reversibility in thermodynamics, or the concept of reversible reactions in kinetics.

In this section we show that
\begin{equation}
 q^-(r)=p_\text{ss}(r)/\pi(r) \label{Langer_q-}
\end{equation}
First, as in the one-dimensional case, the steady-state distribution $p_\text{ss}(r)$ satisfies the following boundary value problem,
\begin{equation}
 \begin{cases}
 L^\dagger p_\text{ss}(r)=0, & r\in\Omega\backslash(A\cup B) \\
 p_\text{ss}(r)=\pi(r), & r\in A \\
 p_\text{ss}(r)=0, & r\in B
 \end{cases}
 \label{Langer_p_ss_eq}
\end{equation}
Second, similar to the MFPT, the backward committor also satisfies a boundary value problem,
\begin{equation}
 \begin{cases}
 Lq^-(r)=0, & r\in\Omega\backslash(A\cup B) \\
 q^{-}(r)=1, & r\in A \\
 q^-(r)=0, & r\in B
 \end{cases}
 \label{Langer_committor_eq}
\end{equation}
In general, the time evolution of the backward committor is actually dictated by $L^R$, the generator for the time-reversed process, but $L^R=L$ for time-reversible processes. Again, we will not attempt to derive \eqref{Langer_committor_eq} rigorously. An ``intuitive'' justification is as follows: consider the same hypothetical procedure as before, where an ensemble of system trajectories are prepared, all with the same initial condition at position $r(0)=r_0$ outside of $A$ or $B$; we can calculate $q^-(r_0)$ by determining the fraction of these trajectories that reach $A$ first before $B$. The fate of each individual realization of the system dynamics is time-independent: a trajectory either eventually reaches $A$ first before it reaches $B$, or it doesn't; therefore, the fraction $q^-(r_0)$ is time-independent, and thus $Lq^-(r_0)$, which gives the time derivative, is zero. If $r_0$ is already in $A$, then $q^-=1$ by construction since it is not possible for the trajectory to reach $B$ before $A$; similarly, if $r_0$ is already in $B$, then $q^-=0$ since the trajectory is already in $B$ before it can reach $A$.

Comparing \eqref{Langer_FP_operator} with \eqref{Langer_generator}, we see that the generator is related to the Fokker-Planck operator by
\begin{equation}
 L=\pi(r)^{-1}L^\dagger\pi(r) \label{generator_FP_operator}
\end{equation}
Substituting \eqref{generator_FP_operator} in \eqref{Langer_committor_eq} gives $\pi(r)^{-1}L^\dagger\pi(r)q^-(r)=0$; this implies that $L^\dagger\pi(r)q^-(r)=0$ since $\pi^{-1}(r)\propto e^{\beta W(r)}$ is strictly positive for all $r$. Furthermore, $\pi(r)q^-(r)=\pi(r)$ for $r\in A$ and $\pi(r)q^-(r)=0$ for $r\in B$. Taken together, we see that $\pi(r)q^-(r)$ satisfies the same well-posed boundary value problem \eqref{Langer_p_ss_eq} as $p_\text{ss}(r)$, which implies that they are the same function, as desired. We note here that this proof does not work for stochastic processes described by the Langevin equation outside the overdamped regime.

\subsection{An ansatz for the backward committor}

Given \eqref{Langer_q-}, the problem of determining $p_\text{ss}(r)$ now reduces to that of determining $q^-(r)$, the backward committor. In this section, we seek an analytical expression for the backward committor near the saddle point $r^\ddagger$. To make the derivation clearer, we first perform a change of variable $s=r-r^\ddagger$ to shift the origin to the saddle point; this procedure has no effect on the functional form of any preceding expressions except for changing the independent variable from $r$ to $s$. At steady state, use the fact that $p_\text{ss}(s)$ is time-independent and $p_\text{ss}(s)=q^-(s)\pi(s)$; \eqref{Langer_FP_elementwise} becomes

\begin{align}
 0 &= \sum_{ij}\partial_iD_{ij}\left[\beta\partial_j W(s)+\partial_j\right]q^-(s)\pi(s) \\
 0 &= \sum_{ij}\partial_i D_{ij}\left[\beta(\partial_jW(s))q^-(s)\pi(s) + (\partial_j q^-(s))\pi(s) - \beta\pi q^-(s)\partial_jW(s)\right] \\
 0 &= \sum_{ij}D_{ij}\partial_i\left[(\partial_j q^-(s))\pi(s)\right] \\
 0 &= \sum_{ij}D_{ij}\left[\partial_{ij} q^-(s) -\beta\left(\partial_iW(s)\right)\left(\partial_j q^-(s)\right)\right] \label{Langer_committor_diff_eq}
\end{align}
In the second equality, we used the fact that $\partial_i\pi(s)=-\beta\pi(s)\partial_iW(s)$.

To examine the solution to \eqref{Langer_committor_diff_eq} near the saddle point $r^\ddagger$, we first consider the one dimensional case. Here, \eqref{Langer_harmonic_saddle} becomes $W(s)=\Delta W^\ddagger - \kappa^\ddagger s^2/2$ as in \eqref{Kramers_harmonic}, with $\kappa^\ddagger>0$ representing the force constant for the unstable mode, and \eqref{Langer_committor_diff_eq} reduces to
\begin{equation}
 \frac{d^2}{ds^2}q^{-}(s) + \beta\kappa^\ddagger s\frac{d}{ds}q^{-}(s)=0 \label{Langer_committor_diff_eq_1D}
\end{equation}
Equation \eqref{Langer_committor_diff_eq_1D} can be solved by first solving a first-order ODE for $\nu(s)=dq^{-}/ds$ using an integrating factor and then integrating both sides of the solution from $0$ to $s$. After some algebra, we arrive at the general solution in the form of an error function
\begin{equation}
 q^-(s)=A+\frac{B}{\sqrt{2\pi}}\int_0^{s\sqrt{\beta\kappa^\ddagger}} e^{-u^2/2}\,du \label{Langer_committor_1D_sol}
\end{equation}
where $A$ and $B$ are two free constants. Since $q^-(s)$ is a backward committor, we require that $q^-\to 1$ as $s\to-\infty$ (i.e., for $s$ in the reactant well) and $q^-\to 0$ as $s\to\infty$ (i.e., for $s$ in the product well). Substituting these limits into \eqref{Langer_committor_1D_sol} gives $A-B/2=1$ and $A+B/2=0$; it should be obvious that $A=1/2$ and $B=-1$, which means that the solution to \eqref{Langer_committor_diff_eq_1D} that satisfies the appropriate boundary conditions is
\begin{equation}
 q^-(s)=\frac 1 2 -\frac{1}{\sqrt{2\pi}}\int_0^{s\sqrt{\beta\kappa^\ddagger}} e^{-u^2/2}\,du = \frac{1}{\sqrt{2\pi}}\int_{s\sqrt{\beta\kappa^\ddagger}}^\infty e^{-u^2/2}\,du = \sqrt{\frac{\beta\kappa^\ddagger}{2\pi}}\int_s^\infty e^{-\beta\kappa^\ddagger u^2/2}\,du \label{Langer_committor_1D_specific_sol}
\end{equation}
This solution indicates that the committor drops down from 1 to 0 with a sigmoidal functional form as $s$ increases. In particular, at $s=0$, $q^-=1/2$.

This analysis of the one-dimensional backward committor provides important hints about the behavior of the committor in higher dimensions. In this setting, the set of all points $s$ for which $q^-(s)$ has the same value constitutes an isocommittor surface (or curve, in two dimensions). Near the transition state, it is not unreasonable to assume that these isocommittor surfaces can be approximated as planes (or lines, in two dimensions) if $q^-$ is sufficiently smooth. Furthermore, within the region where the harmonic approximation is valid, this family of isocommittor planes needs to be parallel, since level surfaces cannot intersect for a well-defined function.

These considerations motivate us to devise a solution to \eqref{Langer_committor_diff_eq} in terms of isocommittor planes. In general, a plane passing through the origin can be parameterized using its normal vector $n$ as $n^Ts=0$; for planes that do not cross the origin, they can still be parameterized as $n^T(s-s_0)=0$, where $s_0$ is a displacement vector from some point on the plane to the origin. In light of this functional form, let us define a vector $v_+$ that is normal to the isocommittor planes (the reason for this notation will become clear shortly); using this vector, the family of isocommittor planes can be parameterized as $v_+^Ts=u$, $u\in\mathbb R$. We orient $v_+$ such that $q^-$ approaches 1 for $u<0$ and approaches 0 for $u>0$.

Taken together, the analysis so far suggests that the solution to \eqref{Langer_committor_diff_eq} has the form
\begin{equation}
 q^-(s)=\frac{a}{\sqrt{2\pi}}\int_{v_+^Ts}^\infty e^{-(au)^2/2}\,du \label{Langer_isocommittor_guess}
\end{equation}
where $a>0$ is a free constant to be determined. It is easy to check that with this solution form, $q^-(s)$ is constant on each isocommittor plane, approaches 1 for $s$ near the reactant well, and approaches 0 for $s$ near to the product well.

One particular isocommittor plane of interest is the one that crosses the saddle point; i.e., the plane parameterized by $u=0$ and where $q^-=1/2$. This plane is sometimes known as the stochastic separatrix and often plays a special role in the analysis of reaction mechanisms: if we define the transition state (or more accurately, the transition state ensemble) as the set of configurations where the committor is 1/2, the transition state is the stochastic saparatrix. In Kramers' theory, this plane reduces to the peak of the barrier separating the reactant and product state. Unfortunately, the simplicity of this equivalence between the transition state and the barrier peak often does not hold in practice, depending on the choice of the reaction coordinate and the theoretical framework under consideration; see, e.g., \citep{vanden-eijndenTransitionStateTheory2005, dellagoTransitionPathSampling2006}.

\subsection{Properties of the vector normal to the isocommittor planes}
\label{v_+_as_eigenvector}

The ansatz in \eqref{Langer_isocommittor_guess} still leaves some questions unanswered. Specifically,
\begin{enumerate}
 \item How does the direction of $v_+$ relate to the geometry at the saddle point?
 \item What is the magnitude of $v_+$?
 \item What is the value of the free constant $a>0$?
\end{enumerate}
We address these questions in this section.

\subsubsection{v\textsubscript{+} as a generalized eigenvector}
To understand the direction of $v_+$, we first argue that $v_+$ is actually an eigenvector of the matrix $H^\ddagger D$. In order to show this, we need to examine the solution of \eqref{Langer_committor_diff_eq} on the $q^-=1/2$ plane.

First, using the Leibniz integral rule, we record here the first- and second-order derivatives of \eqref{Langer_isocommittor_guess} with respect to $s$:
\begin{equation}
 \partial_i q^-(s)=-\frac{a}{\sqrt{2\pi}} e^{-\left(av_+^Ts\right)^2/2}v_{+i} \quad\text{and}\quad \partial_{ij}q^-(s)=\frac{a^3}{\sqrt{2\pi}}\left(v_+^Ts\right) e^{-\left(av_+^Ts\right)^2/2}v_{+i}v_{+j} \label{Langer_committor_derivatives}
\end{equation}
On the $q^-=1/2$ plane, $v_+^Ts=0$ and \eqref{Langer_committor_derivatives} reduces to
\begin{equation}
 \partial_i q^-(s)=-\frac{a}{\sqrt{2\pi}}v_{+i} \quad\text{and}\quad \partial_{ij}q^-(s)=0 \quad\text{when}\quad v_+^Ts=0 \label{Langer_committor_1/2_derivatives}
\end{equation}
Substituting in the derivative of $W(s)$ in \eqref{Langer_quadratic_form_derivative} and the derivatives of $q^-(s)$ in \eqref{Langer_committor_1/2_derivatives}, we see that \eqref{Langer_committor_diff_eq} can be further simplified on the $q^-=1/2$ plane as,
\begin{align}
   0&= \sum_{ij} D_{ij}\left[-\beta\left(\sum_k H^\ddagger_{ik}s_k\right) \left(-\frac{a}{\sqrt{2\pi}}v_{+j}\right)\right]\\
   0&= \sum_k s_k\left[\sum_{ij}H^\ddagger_{ki}D_{ij}v_{+j}\right] \\
   0&= s^T\left(H^\ddagger Dv_+\right)\label{Langer_eigenvector_derivation}
\end{align}
In the second equality, we used the fact that $H^\ddagger$ is symmetric (i.e., $H^\ddagger_{ik}=H^\ddagger_{ki}$). This result indicates that the vector $H^\ddagger D v_+$ is orthogonal to $s$. Since $v_+$ is orthogonal to $s$ confined on the $q^{-1}=1/2$ plane, $H^\ddagger D v_+$ is parallel with $v_+$. This shows that $v_+$ is an eigenvector of $H^\ddagger D$. Anticipating the analysis in the next section, we denote the corresponding eigenvalue as $-\lambda_+$ (with $\lambda_+>0$).

Note that we can write the eigenvector equation $H^\ddagger Dv_+=-\lambda_+v_+$ as
\begin{equation}
H^{-\ddagger}v_+=-\lambda_+^{-1}Dv_+ 
\end{equation}
where, through an abuse of notation, we have written $\left(H^{\ddagger}\right)^{-1}=H^{-\ddagger}$. Equations of this form are called generalized eigenvalue problems whenever $H^{-\ddagger}$ is symmetric and $D$ is symmetric positive definite. The collection of all such generalized eigenvectors form a nonsingular matrix $V$ that simultaneously diagonalizes $H^\ddagger$ and $D$, in the sense that
\begin{equation}
 V^TDV=I \quad\text{and}\quad V^TH^{-\ddagger} V=\Lambda^{-1} \label{gen_eig_diag}
\end{equation}
where $\Lambda^{-1}$ is a diagonal matrix containing the corresponding generalized eigenvalues. The vectors $v_i$'s in $V$ are orthogonal with respect to the inner product induced by $D$; i.e.,
\begin{equation}
 \left<v_i, v_j\right>_D=v_i^TDv_j=\lambda_jv_i^TH^{-\ddagger} v_j=\lambda_j\Lambda_{ij}^{-1}=\delta_{ij}
\end{equation}
where $\delta_{ij}$ is the Kronecker delta function.

\subsubsection{The value of the free constant a}

The magnitudes of $a$ and $v_+$ are related; first we determine the value of $a$ in terms of $v_+$. To do so, we consider the solution to \eqref{Langer_committor_diff_eq} outside the committor 1/2 plane where $v_+^Ts\ne 0$:
\begin{align}
 0 &= \sum_{ij} D_{ij}\left[\frac{a^3}{\sqrt{2\pi}}\left(v_+^Ts\right) e^{-\left(av_+^Ts\right)^2/2}v_{+i}v_{+j} - \beta\left(\sum_k H^\ddagger_{ik}s_k\right) \left(-\frac{a}{\sqrt{2\pi}}e^{-(av_+^Ts)^2/2}v_{+j}\right)\right]\\
 0 &= \sum_{ij} D_{ij}\left[a^2\left(v_+^Ts\right)v_{+i}v_{+j} + \beta\left(\sum_k H^\ddagger_{ik}s_k\right)v_{+j}\right]\\
 0 &= a^2(v_+^Ts)\sum_{ij}v_{+i}D_{ij}v_{+j} + \beta\sum_{jik}v_{+j}D_{ji}H^\ddagger_{ik}s_k\\
 0 &= a^2(v_+^Ts)v_+^TDv_+ + \beta (v_+^TDH^\ddagger)s \label{beta_or_no_beta}\\
 0 &= a^2(v_+^Ts)v_+^TDv_+ - \beta (\lambda_+ v_+^T)s \\
 0 &= a^2v_+^TDv_+ -\beta\lambda_+ \\
 a &= \sqrt{\beta\lambda_+/v_+^TDv_+} \label{free_const_a}
\end{align}
Applying \eqref{free_const_a} to \eqref{Langer_isocommittor_guess} gives
\begin{equation}
 q^-(s)=\sqrt{\frac{\beta\lambda_+}{2\pi v_+^TDv_+}}\int_{v_+^Ts}^\infty \exp\left(-\frac1 2\frac{\beta\lambda^+}{v_+^TDv_+}u^2\right)\,du \label{Langer_isocommittor_guess2}
\end{equation}

Note that \eqref{free_const_a} implies that $-\lambda_+$ is a negative eigenvalue of $H^\ddagger D$: because $\beta>0$ and $v_+^TDv_+>0$ (since $D$ is symmetric positive definite), $\lambda_+$ also needs to be positive so that $a>0$. What does this result mean for the direction represented by the corresponding eigenvector $v_+$? When $D$ can be written in the form of $\sigma^2 I$, where $I$ is an identity matrix, the diffusion on the potential (of mean force) is isotropic (i.e., same diffusivity in every direction). In this case, $v_+$ is simply an eigenvector of $H^\ddagger$,
\begin{equation}
 H^\ddagger v_+=-\lambda_+\sigma^{-2} v_+
\end{equation}
that is, $v_+$ is an eigenvector associated with the only negative eigenvalue of $H^\ddagger$ corresponding to the unstable mode separating the reactant and product well. In the case of anisotropic diffusion, $H^\ddagger D$ still has a single unique unstable mode represented by $v_+$: because $H^\ddagger$ is symmetric and $D$ is symmetric positive definite, a result in linear algebra states that the eigenvalues of $H^\ddagger D$ have the same signs as those of $H^\ddagger$. As such, $v_+$ is still an eigenvector associated with the only negative eigenvalue of $H^\ddagger D$ in the anisotropic case. However, the presence of the matrix $D$ effectively stretches and rotates the dynamics on the surface of the potential; as such, the unstable mode of $H^\ddagger D$ no longer necessarily points in the same direction as the unstable mode of $H^\ddagger$ (see, e.g., Figure \ref{fig:2d_example}A).  

\subsubsection{The magnitude of v\textsubscript{+}}

The magnitude of $v_+$ determines how fast the committor $q^-(s)$ decays to zero from the reactant well to the product well. To determine this quantity, we first note that the derivative of the one-dimensional backward committor in \eqref{Langer_committor_1D_specific_sol} at the saddle point $s=0$ is given by
\begin{equation}
 \frac{d}{ds}q^-(s)\Big|_{s=0}=-\sqrt{\frac{\beta\kappa^\ddagger}{2\pi}}e^{-\beta\kappa^\ddagger s^2/2}\Big|_{s=0}=-\sqrt{\frac{\beta\kappa^\ddagger}{2\pi}} \label{committor_1D_derivative}
\end{equation}
While the gradient of the multi-dimensional backward committor in \eqref{Langer_isocommittor_guess2} at the saddle point $s=0$ is given by
\begin{equation}
 \nabla_s q^-(s)\Big|_{s=0} = -\sqrt{\frac{\beta\lambda_+}{2\pi v_+^TDv_+}}\exp\left(-\frac1 2\frac{\beta\lambda^+}{v_+^TDv_+}(v_+^Ts)^2\right)\Big|_{s=0} v_+=-\sqrt{\frac{\beta\lambda_+}{2\pi v_+^TDv_+}}v_+ \label{committor_ND_derivative}
\end{equation}
Comparing these two expressions, we can identify $\kappa^\ddagger$ with $\lambda_+$ since both are ``eigenvalues'' corresponding to the unstable mode at the saddle point. In order for the absolute value of \eqref{committor_1D_derivative} and the norm of \eqref{committor_ND_derivative} (induced by $D$) to be equal, we require that
\begin{equation}
 v_+^TDv_+=1
\end{equation}
Recall that the expression $v_+^TDv_+$ is the inner product of $v_+$ with itself induced by $D$ and defines a vector norm via $\|v_+\|_D=\sqrt{v_+^TDv_+}$. Taken together, we see that the solution to \eqref{Langer_committor_diff_eq} is
\begin{equation}
  q^-(s)=\sqrt{\frac{\beta\lambda_+}{2\pi}}\int_{v_+^Ts}^\infty e^{-\beta\lambda_+u^2/2}\,du \label{Langer_isocommittor_solution}
\end{equation}
See Figure \ref{fig:2d_example}B and \ref{fig:2d_example}C for a comparison of \eqref{Langer_isocommittor_solution} with the exact solution to \eqref{Langer_committor_eq} for the example double-well potential shown in \ref{fig:2d_example}A.

Lastly, we note here that, with the help of \eqref{Langer_committor_derivatives} (after setting $a=\sqrt{\beta\lambda_+}$), it is fairly straightforward to check that \eqref{Langer_isocommittor_solution} is in fact an exact solution to \eqref{Langer_committor_diff_eq} under the harmonic approximation; i.e., \eqref{Langer_isocommittor_solution} satisfies
\begin{equation}
 \sum_{ij}D_{ij}\partial_{ij}q^-(s)-\beta\sum_{ijk}s_kH^\ddagger_{ki}D_{ij}\partial_j q^-(s)=0
\end{equation}
in addition to satisfying the boundary conditions listed in \eqref{Langer_committor_eq}. In other words, if the potential surface $W(s)$ is exactly harmonic, the isocommittor surfaces are indeed parallel planes parametrized by $v_+^Ts=u$ for $u\in\mathbb R$.

\subsection{Properties of the steady-state flux vector field}
\label{properties_of_J(s)}

Now with an expression for the backward committor in hand, we are ready to write down an expression for the steady-state flux and analyze its behavior near the saddle point.

\subsubsection{The flux vector field has a constant direction}

Substituting \eqref{Langer_q-} and \eqref{Langer_isocommittor_solution} in the multidimensional Fokker-Planck equation \eqref{Langer_FP}, we see that the steady-state flux is
\begin{align}
 J(s) &= -D\pi(s)\nabla_s q^-(s) \label{Langer_ss_flux_def}\\
   &= -D\pi(s)\left(-\sqrt{\frac{\beta\lambda_+}{2\pi}}e^{-\beta\lambda_+(v_+^Ts)^2/2}\right)\nabla_s(v_+^Ts) \\
   &= \sqrt{\frac{\beta\lambda_+}{2\pi}}e^{-\beta\lambda_+(v_+^Ts)^2/2}\pi(s) D v_+ \label{Langer_ss_flux}
\end{align}
One immediate observation here is that near the saddle point, the flux has a constant direction $Dv_+$.

\subsubsection{The flux vector field is constant along its streamlines}

Not only does $J(s)$ have a constant direction $Dv_+$, it also has a constant magnitude along that direction. To see why, we first parameterize $s$ using an orthogonal decomposition
\begin{equation}
 s(u)=\Delta s+uDv_+
\end{equation}
where $\Delta s$ is a constant that measures the displacement from the origin (i.e., the saddle point) orthogonal to the direction defined by $Dv_+$, while $u\in\mathbb R$ is the independent variable that measures progression along the $Dv_+$ direction. In \eqref{Langer_ss_flux}, the only factor dependent on $s$ is $e^{-\beta\lambda_+(v_+^Ts)^2/2}\pi(s)$. Using the harmonic approximation of $\pi(s)$ at near the saddle point in \eqref{Langer_harmonic_saddle}, this expression can be written as
\begin{equation}
 \exp\left[-\frac\beta 2 s^T\left(H^\ddagger +\lambda_+v_+v_+^T\right)s\right]
\end{equation}
Using the parameterization $s(u)$, the quadratic form in this expression becomes $(\Delta s+uDv_+)^T(H^\ddagger +\lambda_+v_+v_+^T)(\Delta s+uDv_+)$. This expression can be expanded, but some of the terms after the expansion are constants independent of $u$. Keeping only the $u$-dependent terms, we get
\begin{multline}
 u\Delta s^TH^\ddagger Dv_+ + u\lambda_+\Delta s^Tv_+v_+^TDv_+ + u v_+^TD^TH^\ddagger \Delta s \\ + u^2v_+^TD^TH^\ddagger Dv_+ + u\lambda_+v_+^TD^Tv_+v_+^T\Delta s + u^2\lambda_+v_+^TD^Tv_+v_+^TDv_+
\end{multline}
Using the facts that $H^\ddagger$ and $D$ are symmetric, $H^\ddagger Dv_+=-\lambda_+v_+$, and $v_+^TDv_+=1$, one can show that every term in this expression cancels, and thus $J(s)$ is constant along the $Dv_+$ direction. Taken together, we can visualize the flux as a flow of probability density along parallel streamlines near the saddle point, and the velocity of the probability current along each streamline is constant. See Figure \ref{fig:2d_example}B and \ref{fig:2d_example}C for a comparison of the streamlines of \eqref{Langer_ss_flux} with the exact solution to \eqref{Langer_ss_flux_def} for the example double-well potential shown in \ref{fig:2d_example}A.

\subsubsection{Flux surface integrals are constant over dividing surfaces}
\label{flux_surface_integral}

As in the one-dimensional case, we need the total flux from the reactant well to the product well, which gives the reaction rate. This task is made somewhat easy by a few considerations. First, we note that, for a double-well potential, the flux-over-population steady-state flux $J(s)$ has exactly one source in the reactant well $A$ and one sink in the product well $B$; that is, $J(s)$ is a divergence-less vector field in $\Omega\backslash(A\cup B)$ at the steady state (this should be obvious by setting time derivative to zero in \eqref{Langer_FP}). A consequence of this property is that the total flux from $A$ to $B$ is given by the surface integral of $J(s)$ over a dividing surface between $A$ and $B$ (i.e., a surface that partitions $\Omega$ into two sets, one of which contains $A$ and the other contains $B$). Importantly, the particular choice of the dividing surface is irrelevant; if this integral differs on two dividing surfaces, it would imply that there is a net sink or source located in the in-between region. Second, since the probability flow is concentrated near the saddle point, one can restrict the flux surface integral further to regions close to the saddle point where \eqref{Langer_ss_flux} is valid.

With the help of \eqref{gaussian_delta_integral}, a result we will prove in Section \ref{BS_simplification}, a particularly simple choice of dividing surface is the $q^-=1/2$ plane. The total flux of \eqref{Langer_ss_flux} over the stochastic separatrix is
\begin{align}
 J &= \int \|v_+\|^{-1}v_+^TJ(s)\,d\sigma(s) \\
   &\overset{1}{=} \int v_+^TJ(s)\delta(v_+^Ts)\,ds \\
   &= \sqrt{\frac{\beta\lambda_+}{2\pi}}v_+^TD v_+ Z^{-1}e^{-\beta\Delta W^\ddagger} \int e^{^-\beta s^TH^\ddagger s/2}\delta(v_+^Ts)\,ds \\
   &= \sqrt{\frac{\beta\lambda_+}{2\pi}}Z^{-1}e^{-\beta\Delta W^\ddagger} \left[(2\pi/\beta)^{1-N} v_+^TH^{-\ddagger}v_+\det H^\ddagger\right]^{-1/2} \\
   &\overset{2}{=} \left(\frac{2\pi}{\beta}\right)^{N/2}Z^{-1}e^{-\beta\Delta W^\ddagger}\frac{\beta\lambda_+}{2\pi}\left|\det H^\ddagger\right|^{-1/2} \label{Langer_flux_integral_sol_q_1/2}
\end{align}
Here, equality (1) follows from the relation $d\sigma(s)=\delta(v_+^Ts)\|\nabla_s v_+^Ts\|\,ds$, which is a consequence of what is known as the coarea formula; note that $\|\nabla_s v_+^Ts\|=\|v_+\|$. The volume integrals are over the entire $\Omega=\mathbb R^N$ reaction coordinate space of the system; this will be the assumed domain of integration for the rest of the note unless otherwise specified. Equality (2) follows from the facts that $H^{-\ddagger}v_+=-\lambda_+^{-1}Dv_+$ and $\det H^\ddagger<0$.

In the rest of this section, we show and confirm that the total flux can be obtained using a surface integral over an arbitrary plane near the the saddle point, which gives the same result as \eqref{Langer_flux_integral_sol_q_1/2}. Let us parameterize such a plane by $n^Ts=\theta$, where $n$ is the normal vector (not necessarily normalized) and $\theta\in\mathbb R$. The only condition we impose on the plane is that $n$ is not orthogonal to $Dv_+$ (i.e., the plane parameterized by $n^Ts=\theta$ is not parallel to $Dv_+$, the direction of the flux). Furthermore, without loss of generality we assume that $n$ is oriented such that $n^TDv_+>0$. Using \eqref{Langer_ss_flux} and the coarea formula, the integral of $J(s)$ over this plane is given by
\begin{align}
 J &= \int \|n\|^{-1}n^TJ(s)\,d\sigma(s) \\
   &= \int n^TJ(s)\delta(\theta-n^Ts)\,ds \\
   &= \sqrt{\frac{\beta\lambda_+}{2\pi}}n^TD v_+ Z^{-1}e^{-\beta\Delta W^\ddagger} \int e^{^-\beta s^T\left(H^\ddagger+\lambda_+v_+v_+^T\right) s/2}\delta(\theta - n^Ts)\,ds \label{Langer_flux_integral}
\end{align}

As with most Gaussian integrals, we seek a change of variable that makes the volume integral in \eqref{Langer_flux_integral} separable. For a typical Gaussian integral with an integrand of the form $e^{-x^TAx/2}$, where $A$ is a symmetric matrix, there is an orthogonal matrix $Q$ that diagonalizes $A$ (i.e., $Q^TAQ=M$ for some diagonal matrix $M$) and enables a change of variable $x=Qy$ that renders the integral separable. In the current case, recall our discussion of the generalized eigenvector problem in Section \ref{v_+_as_eigenvector}, where we have defined the matrix $V$ that simultaneously diagonalizes $H^\ddagger$ and $D$ according to \eqref{gen_eig_diag}. Comparing $V$ and $Q$, we introduce the change of variable $t=V^Ts$ (as well as $m=V^Tn$). The Jacobian for this transformation is $\det V^{-1}$, which is not necessarily one because $V$ may not be orthogonal. After this transformation, $v_+$ relates to a new basis vector $e_+$ by $e_+=V^{-1}v_+$; to see why, note that together with \eqref{gen_eig_diag}, $H^\ddagger Dv_+=-\lambda_+v_+$ implies
\begin{equation}
 \Lambda V^{-1}v_+=-\lambda_+V^{-1}v_+ \label{norm_vector_basis}
\end{equation}
that is, $V^{-1}v_+$ is an eigenvector of the diagonal matrix $\Lambda$ corresponding to its unique negative eigenvalue $-\lambda_+$. Since $-\lambda_+$ has an algebraic multiplicity of one, the eigenvector $V^{-1}v_+$ must have only one nonzero element. Furthermore, the norm of $V^{-1}v_+$ is fixed by $\|V^{-1}v_+\|=v_+V^{-T}V^{-1}v_+=v_+^TDv_+=1$. Since $v_+$ is a column vector of $V$, the nonzero element of $e_+$ must be positive, otherwise $Ve_+=-v_+$. Together, we have shown that $e_+$ is a normalized standard basis vector in the new coordinate system.

With this change of variable, \eqref{Langer_flux_integral} becomes
\begin{align}
 J 
 &= \sqrt{\frac{\beta\lambda_+}{2\pi}}n^TD v_+ Z^{-1}e^{-\beta\Delta W^\ddagger} \int e^{^-\beta t^TV^{-1}\left(H^\ddagger+\lambda_+v_+v_+^T\right)V^{-T}t/2}\delta(\theta - m^TV^{-1}V^{-T}t)|\det V^{-1}|\,dt\\
 &= \sqrt{\frac{\beta\lambda_+}{2\pi}}n^TD v_+ Z^{-1}e^{-\beta\Delta W^\ddagger}|\det V^{-1}| \int e^{^-\beta t^T\left(\Lambda+\lambda_+e_+e_+^T\right)t/2}\delta(\theta - m^TDt)\,dt \label{Langer_flux_integral_2}
\end{align}
Before further simplifications of \eqref{Langer_flux_integral_2}, a comment about the term $t^T\left(\Lambda+\lambda_+e_+e_+^T\right)t$ is in order. In the new basis, $H^\ddagger$ becomes $\Lambda$, while $v_+v_+^T$ becomes $e_+e_+^T$. In particular, $\Lambda+\lambda_+e_+e_+^T$ is a diagonal matrix whose diagonal elements are the eigenvalues $\lambda_j$'s except for $-\lambda_+$, which has been deleted from $\Lambda$ by $\lambda_+e_+e_+^T$. This observation should become especially obvious using the outer product form of the spectral theorem, which gives $\Lambda=\sum_j\lambda_je_je_j^T$; in other words, compared to $\Lambda$ (or $H^\ddagger$), $\Lambda+\lambda_+e_+e_+^T$ (or $H^\ddagger+\lambda_+v_+v_+^T$) is rank-deficient because one of its eigenspaces corresponding to the unstable mode has been deleted. As a result, the quadratic form $t^T(\Lambda+\lambda_+e_+e_+^T)t$ sums over all $\lambda_jt_j^2$ except for $j$ corresponding to $-\lambda_+t_+^2$. In light of this analysis, we will use notations $\sum_{j\ne+}$ and $\prod_{j\ne+}$ to denote summation or product over $j$'s except for $j$ corresponding to $-\lambda_+$ and $e_+$.

Let us denote the volume integral in \eqref{Langer_flux_integral_2} as $I$. We will simplify $I$ with the following strategy. First, we convert the Dirac delta function in \eqref{Langer_flux_integral_2} into its integral representation; i.e.,
\begin{equation}
 \delta(f(x))=\frac{1}{2\pi}\int_{-\infty}^\infty e^{ikf(x)}\,dk \label{dirac_delta_integral}
\end{equation}
for any sufficiently smooth test function $f(x)$. Second, we integrate over all $t_j$'s for which $j\ne+$, then over $k$, and then over $t_+$. This order of operation ensures that every integral is an analytically tractable Gaussian integral with a linear term, which can be solved by first completing the squares:
\begin{align}
 I&= \frac{1}{2\pi}\int\int e^{^-\beta t^T\left(\Lambda+\lambda_+e_+e_+^T\right)t/2}e^{ik(\theta - m^TDt)}\,dtdk\\
 &= \frac{1}{2\pi}\int\int e^{ik\theta}e^{-ik\sum_l m_lD_{l+}t_+}\prod_{j\ne+}\left(\int e^{^-\beta \lambda_jt_j^2/2}e^{-ik\sum_lm_lD_{lj}t_j}\,dt_j\right)\,dkdt_+\\
 &= \frac{1}{2\pi}\int\int e^{ik\theta}e^{-ik\sum_l m_lD_{l+}t_+}\prod_{j\ne+}\left(\sqrt{\frac{2\pi}{\beta\lambda_j}}e^{-k^2\left(\sum_l m_lD_{lj}\right)^2/2\beta\lambda_j}\right)\,dkdt_+\\
 &= \frac{1}{2\pi}\left(\prod_{j\ne+}\sqrt{\frac{2\pi}{\beta\lambda_j}}\right)\int e^{ik\theta}e^{-ik\sum_l m_lD_{l+}t_+}\left(\frac{2\pi\beta}{\sum_{j\ne+}\lambda_j^{-1}\left(\sum_l m_lD_{lj}\right)^2}\right)^{1/2}e^{-\frac{\beta}{2}\frac{(\theta-\sum_l m_lD_{l+}t_+)^2}{\sum_{j\ne+}\lambda_j^{-1}\left(\sum_l m_lD_{lj}\right)^2}}\,dt_+\\
&= \frac{1}{2\pi}\left(\prod_{j\ne+}\sqrt{\frac{2\pi}{\beta\lambda_j}}\right)\left(\frac{2\pi\beta}{\sum_{j\ne+}\lambda_j^{-1}\left(\sum_l m_lD_{lj}\right)^2}\right)^{1/2}\left(\frac{2\pi}{\beta}\frac{\sum_{j\ne+}\lambda_j^{-1}\left(\sum_l m_lD_{lj}\right)^2}{\left(\sum_l m_lD_{l+}\right)^2}\right)^{1/2}\\
&= \left(\prod_{j\ne+}\sqrt{\frac{2\pi}{\beta\lambda_j}}\right)|m^TDe_+|^{-1} \label{Langer_flux_integral_3}
\end{align}
Substituting \eqref{Langer_flux_integral_3} back into \eqref{Langer_flux_integral_2} gives
\begin{align}
 J &= \sqrt{\frac{\beta\lambda_+}{2\pi}}\left(\prod_{j\ne+}\sqrt{\frac{2\pi}{\beta\lambda_j}}\right) Z^{-1}e^{-\beta\Delta W^\ddagger}|\det V^{-1}|\frac{n^TD v_+}{|m^TDe_+|}\\
 &\overset{1}{=} \left(\frac{2\pi}{\beta}\right)^{N/2}\frac{\beta \lambda_+}{2\pi}Z^{-1}e^{-\beta\Delta W^\ddagger}\frac{|\det V^{-1}|}{|\det H^\ddagger D|^{1/2}} \\
 &\overset{2}{=} \left(\frac{2\pi}{\beta}\right)^{N/2}Z^{-1}e^{-\beta\Delta W^\ddagger}\frac{\beta \lambda_+}{2\pi}|\det H^\ddagger|^{-1/2} \label{Langer_flux_integral_sol}
\end{align}
In equality (1), we have used the fact that
\begin{equation}
 n^TDv_+=m^TV^{-1}DVe_+=m^TV^{-1}\left(V^{-T}V^{-1}\right)Ve_+=m^T\left(V^{-1}V^{-T}\right)e_+=m^TD^Te_+=m^TDe_+>0
\end{equation}
In equality (2), we have used the fact that
\begin{equation}
 |\det H^\ddagger D|=|\det H^\ddagger|\det D \quad\text{and}\quad \det D=\det V^{-T}V^{-1} =\left(\det V^{-1}\right)^2
\end{equation}
both of which follow directly from \eqref{gen_eig_diag}. It should be obvious that \eqref{Langer_flux_integral_sol} is independent of either $\theta$ or $n$ used to parameterize the dividing plane and is thus an intrinsic property of the flux. The result in \eqref{Langer_flux_integral_sol} is the same as \eqref{Langer_flux_integral_sol_q_1/2}, as expected. However, the choice of $q^-=1/2$ in the derivation of \eqref{Langer_flux_integral_sol_q_1/2} is special in that the quadratic form $s^T(H^\ddagger+\lambda_+v_+v_+^T)s$ is simply $s^TH^\ddagger s$, which is not rank-deficient.

\subsection{Langer's rate constant}

According to the flux-over-population method, the only quantity left to determine for computing the rate constant is the steady-state population $n$ in the reactant well. This is given by
\begin{multline}
 n=Z^{-1}\int e^{-\beta (r-r_A)^TH_A (r-r_A)/2}\,dr = Z^{-1}\prod_i \int_{-\infty}^\infty e^{-\beta \mu_i t_i^2/2}\,dt_i \\ =Z^{-1}\left(2\pi/\beta\right)^{N/2}\prod_i\mu_i^{-1/2}=Z^{-1}\left(2\pi/\beta\right)^{N/2}\left(\det H_A\right)^{-1/2} \label{Langer_ss_pop}
\end{multline}
where $r-r_A=Qt$ with $Q$ being an orthogonal matrix that diagonalizes $H_A$, and $\mu_i$'s are the eigenvalues of $H_A$. Recall that $H_A$ is assumed to be symmetric positive definite, and thus $\mu_i>0$ for all $i$'s.

At last, combining \eqref{Langer_flux_integral_sol_q_1/2} and \eqref{Langer_ss_pop}, we see that the rate constant is
\begin{equation}
 k_{AB}=J/n=\frac{\beta\lambda_+}{2\pi}\sqrt{\frac{\det H_A}{\left|\det H^\ddagger\right|}}e^{-\beta\Delta W^\ddagger} \quad\text{where}\quad H^\ddagger Dv_+=-\lambda_+v_+  \label{Langer_rate_constant}
\end{equation}
Three brief comments about this result are in order. First, we note here that it is also common in the literature to define $-\lambda_+$ as an eigenvalue of the matrix $\beta H^\ddagger D$, while the magnitude of $v_+$ is still fixed by $v_+^TDv_+=1$. One can retrace our derivations and see that, starting from \eqref{beta_or_no_beta}, this has the effect of replacing all occurrences of $\beta\lambda_+$ by $\lambda_+$, in which case the rate constant now reads
\begin{equation}
 k_{AB}=\frac{\lambda_+}{2\pi}\sqrt{\frac{\det H_A}{\left|\det H^\ddagger\right|}}e^{-\beta\Delta W^\ddagger} \quad\text{where}\quad \beta H^\ddagger Dv_+=-\lambda_+v_+
\end{equation}

Second, it is easy to check that the multidimensional result can be reduced to Kramers' rate constant \eqref{Kramers_rate_2} in the case of a single reaction coordinate. By comparing \eqref{Kramers_harmonic} to \eqref{Langer_harmonic_reactant} and \eqref{Langer_harmonic_saddle}, we identify $H_A$ as $\kappa_A$ and $H^\ddagger$ as $-\kappa^\ddagger$. Next, since $-\lambda_+$ is the eigenvalue corresponding to the unstable mode of $H^\ddagger D$, and the eigenvalue of a $1\times 1$ ``matrix'' is the matrix element itself, it follows that in the one-dimensional case where only the unstable mode is considered, $-\lambda_+$ can be identified as the product of $H^\ddagger$ (i.e., $-\kappa^\ddagger$) and $D$, the one-dimensional diffusion constant. Making these substitutions reduce \eqref{Langer_rate_constant} to \eqref{Kramers_rate_2}.

Lastly, we briefly comment on some computational aspects of the theory. Using molecular dynamics simulations, the application of \eqref{Langer_rate_constant} requires the determination of the activation free energy $\Delta W^\ddagger$ and the diffusion matrix $D$. The activation free energy can be obtained through a variety of well-established equilibrium techniques such as umbrella sampling, or more recent nonequilibrium techniques based on Jarzynski equality \citep{dellagoComputingEquilibriumFree2014}. Some methods for computing the diffusion matrix are described in, e.g., \citep{imIonsCounterionsBiological2002, liuCalculationDiffusionCoefficients2004, hummerPositiondependentDiffusionCoefficients2005, ma_dynamic_2006, petersCompetingNucleationPathways2009}.

\subsection{Limitation of Langer's theory}

Some caution should be exercised in applying Langer's theory; in this section we briefly discuss two such pitfalls. First, Langer's result can fail in some unexpected cases where the assumption of separation of timescales implied by the flux-over-population procedure is violated. In the case of highly anisotropic diffusion, it is possible to arrange the relative positions of the reactant and product wells such that barrier (re)crossing over the saddle point can take place on a timescale much shorter than the time required to relax within either well due to slower within-well diffusion. If the diffusion anisotropy is extreme, the overall kinetics could become nonexponential, depending on the initial preparation of the system within the reactant well. In such cases, Langer's theory predicts a rate constant that cannot properly account for the dynamics of within-well equilibration. This issue was first described in \citep{berezhkovskiiAnomalousRegimeDecay1989}, and a corrected analytical expression for the rate constant in the two-dimensional case was derived in \citep{berezhkovskiiMultidimensionalReactionRate2014}. More generally, anisotropy in the diffusion matrix and/or potential surface can lead to deviations from Langer's theory; some case studies have been documented in, e.g., \citep{northrupSaddlePointAvoidance1983, klosek-dygasDiffusionTheoryMultidimensional1989}.

Second, for complex systems typically encountered in condensed-phase chemistry and biophysics, the difficulty with Langer's theory is not the calculations entailed by the theory itself, but rather the question of how to select a small number of reaction coordinates that can provide a ``complete'' description of an activated rate process. This question is an active area of research (see, e.g., \citep{chenMolecularEnhancedSampling2018a, ribeiroReweightedAutoencodedVariational2018}), and in practice answering this question is often more akin to an art that relies on domain-specific knowledge and intuition. Typically, an incomplete description of the reaction leads to a loss of Markovianity of the projected stochastic process in the reaction coordinate space; mathematically, this prevents us from simplifying the generalized Langevin equation into the Langevin equation (see Section \ref{review_langevin}), which was the starting point for the derivation of Kramers' theory. When the memory kernel can be reasonably approximated, one approach to treating the non-Markovian dynamics in the reaction coordinate space is the Grote-Hynes theory based on the stable states picture \citep{groteStableStatesPicture1980}, which is akin to Kramers' theory but with the generalized Langevin equation as its starting point; a multidimensional generalization was developed in \citep{berezhkovskiiActivatedRateProcesses1992}. Recent theoretical development has focused on alternative frameworks such as transition path theory \citep{metznerTransitionPathTheory2009}, milestoning \citep{bello-rivasExactMilestoning2015}, weighted ensemble \citep{zuckermanWeightedEnsembleSimulation2017}, and path sampling methods \citep{dellagoTransitionPathSampling1998, vanerpElaboratingTransitionInterface2005} that circumvent these difficulties to various extent, usually at the expense of increased computational costs.

\section{Simplification of Langer's theory by Berezhkovskii and Szabo}
\label{BS_simplification}

It is often claimed that the committor is the ideal reaction coordinate to describe a reaction. In cases where Langer's theory is adequate, this claim is supported by a striking result \citep{berezhkovskii_one-dimensional_2004}, which we will demonstrate in this section; namely, that projecting the system dynamics along the direction of $v_+$ and then applying Kramers' theory gives the same rate constant as Langer's result, while projections along any other directions either give an overestimation or outright do not converge. Recall from our extensive discussion in Section \ref{Langer_theory} that $v_+$ is the vector normal to the isocommittor planes near the saddle point and that $v_+$ is the eigenvector of $H^\ddagger D$ corresponding to the unique unstable mode at the saddle point.

\subsection{Gaussian surface integral over a plane}

Before we discuss the main result of this section, we take a detour here and show that for any vectors $x, x_0, n\in \mathbb R^N$ and an invertible, symmetric matrix $A\in\mathbb R^{N\times N}$,
\begin{equation}
 \int_\Omega\delta(\theta-n^Tx)e^{-\beta(x-x_0)^TA(x-x_0)/2}\,dx=\frac{e^{-\beta(\theta-n^Tx_0)^2/2n^TA^{-1}n}}{\sqrt{(2\pi/\beta)^{1-N}(n^TA^{-1}n)\det A}} \label{gaussian_delta_integral}
\end{equation}
whenever $n^TA^{-1}n\det A>0$. This is a Gaussian integral restricted to a plane parameterized by $n^Tx=\theta$, with $n$ being the normal vector.

Our strategy for evaluating \eqref{gaussian_delta_integral} is somewhat similar to the approach taken in Section \ref{properties_of_J(s)} for evaluating the surface integral of the steady-state flux vector; one major difference here is that \eqref{gaussian_delta_integral} does not involve a rank-deficient matrix. Again, we start the derivation by seeking a change of variable that makes the volume integral in \eqref{gaussian_delta_integral} separable. Since $A$ is symmetric, there exists a diagonal matrix $\Lambda$ and orthogonal matrix $Q$ such that $Q^TAQ=\Lambda$. Here, the diagonal elements in $\Lambda$ contain the eigenvalues of $A$ and the columns of $Q$ are the corresponding eigenvectors. Let us define the change of variable $x=Qy$, along with $x_0=Qy_0$ and $n= Qm$. The Jacobian of this transformation is $\det Q=1$.

Before we apply the change of variable to \eqref{gaussian_delta_integral}, let us pause for a moment and consider the meaning of the condition $n^TA^{-1}n\det A>0$. Since $A$ is invertible and symmetric, all the eigenvalues of $A$ are real and nonzero. If $A$ is furthermore positive definite, then $\det A=\prod\lambda_j>0$ and $n^TA^{-1}n>0$ and thus \eqref{gaussian_delta_integral} converges for any nonzero $n\in\mathbb R^N$. We are more interested in the scenario where $A$ has a single unstable mode. Let us denote this eigenvector as $v_{j'}$ and the corresponding negative eigenvalue as $\lambda_{j'}$. In this case, $\det A<0$ and thus we require $n^TA^{-1}n<0$; this implies that $n$ should not be too close to being perpendicular to $v_{j'}$ (i.e., the integration should not be over a plane close to being parallel to the unstable mode). To see why, let us write $n=\sum_j m_jv_j$ using the eigenvectors $v_j$'s in $Q$ as the basis vectors. With this representation, $n^TA^{-1}n=\sum_j\lambda_j^{-1}m_j^2$; this sum is negative only if $m_{j'}$ is large enough such that $|\lambda_{j'}^{-1}|m_{j'}^2>\sum_{j\ne j'}\lambda_j^{-1}m_j^2$. In the following analysis we will prove \eqref{gaussian_delta_integral} in the case of a single unstable mode.  For readers not interested in the algebraic details, the rest of the section can be skipped without loss of continuity.

Now, let us denote the integral in \eqref{gaussian_delta_integral} as $I$. With the change of variable, we see that
\begin{align}
 I &= \int\delta(\theta-m^TQ^TQy)e^{-\beta (y-y_0)^T\Lambda (y-y_0)/2}\,dy\\
 &= \frac{1}{2\pi}e^{-\beta y_0^T\Lambda y_0/2}\int\int e^{ik(\theta-m^Ty)}e^{-\beta y^T\Lambda y/2}e^{\beta y_0^T\Lambda y}\,dydk\\
 &= \frac{1}{2\pi}e^{-\beta y_0^T\Lambda y_0/2}\int e^{ik\theta}\int e^{-\beta y^T\Lambda y/2}e^{(\beta y_0-ik\Lambda^{-1}m)^T\Lambda y}\,dydk \\
 &= \frac{1}{2\pi}e^{-\beta y_0^T\Lambda y_0/2}\int e^{ik\theta} \prod_j\int e^{-\frac{\lambda_{j}}{2\beta}\left[\beta(y_j-y_{0j}) + ik\lambda_{j}^{-1}m_j\right]^2 + \frac{\lambda_{j}}{2\beta}\left[ik\lambda_{j}^{-1}m_j-\beta y_{0j}\right]^2}\,dy_jdk \\
 &= \frac{1}{2\pi}e^{-\beta y_0^T\Lambda y_0/2}\int e^{ik\theta} \prod_je^{\frac{\lambda_{j}}{2\beta}\left[ik\lambda_{j}^{-1}m_j-\beta y_{0j}\right]^2} \left(\int e^{-\frac{\lambda_{j}}{2\beta}\left[\beta(y_j-y_{0j}) + ik\lambda_{j}^{-1}m_j\right]^2}\,dy_j\right)\,dk \label{all_gaussian_integrals}
\end{align}
The second equality follows by using the integral representation of the Dirac delta function as in \eqref{dirac_delta_integral}.

At this point, we split the integrand of $\int dk$ in \eqref{all_gaussian_integrals} into the product of two expressions. The first involves all $j$'s for which $\lambda_j>0$, and the second consists of $j'$ for which $\lambda_{j'}<0$. The first group is
\begin{multline}
 \prod_{j\ne j'}e^{\frac{\lambda_{j}}{2\beta}\left[ik\lambda_{j}^{-1}m_j-\beta y_{0j}\right]^2}\int e^{-\frac{\lambda_{j}}{2\beta}\left[\beta(y_j-y_{0j}) + ik\lambda_{j}^{-1}m_j\right]^2}\,dy_j = \prod_{j\ne j'}e^{\frac{\lambda_{j}}{2\beta}\left[ik\lambda_{j}^{-1}m_j-\beta y_{0j}\right]^2}\sqrt{\frac{2\pi}{\beta\lambda_j}}\\
 = \left(\prod_{j\ne j'}\sqrt{\frac{2\pi}{\beta\lambda_{j}}}\right) e^{\frac{\beta}{2}\sum_{j\ne j'}\lambda_jy_{0j}^2}e^{\sum_{j\ne j'}\left(-k^2\lambda_j^{-1}m_j^2/2\beta-ikm_jy_{0j}\right)}
\end{multline}
While the second group is simply
\begin{equation}
 \int e^{ik\theta - ikm_{j'}y_{j'}-\beta\lambda_{j'}y_{j'}^2/2+\beta\lambda_{j'}y_{0j'}y_{j'}}\,dy_{j'}
\end{equation}
The original integral in \eqref{gaussian_delta_integral} can now be rewritten as
\begin{align}
 I &= \frac{1}{2\pi}e^{-\frac{\beta}{2} y_0^T\Lambda y_0}\left(\prod_{j\ne j'}\sqrt{\frac{2\pi}{\beta\lambda_{j}}}\right) e^{\frac{\beta}{2}\sum_{j\ne j'}\lambda_jy_{0j}^2}\int \int I(k)I(y_{j'})\,dy_{j'}dk\\
 &= \frac{1}{2\pi}e^{-\frac{\beta}{2}\lambda_{j'}y_{0j'}^2}(2\pi/\beta)^{N-1}\left(\prod_{j\ne j'}\lambda_{j}^{-1/2}\right)\int I(y_{j'})\int I(k)\,dkdy_{j'}
\end{align}
with
\begin{align}
 I(k) &= e^{ik\theta - ikm_{j'}y_{j'} + \sum_{j\ne j'}\left(-k^2\lambda_j^{-1}m_j^2/2\beta-ikm_jy_{0j}\right)} \\
 I(y_{j'}) &= e^{-\frac{\beta}{2}\lambda_{j'}y_{j'}^2+\beta\lambda_{j'}y_{0j'}y_{j'}}
\end{align}
To further simplify this expression, we first integrate over $k$, which is a Gaussian integral with a linear term,
\begin{equation}
 \int I(k)\,dk = \left(\frac{2\pi\beta}{\sum_{j\ne j'}\lambda_j^{-1}m_j^2}\right)^{1/2} e^{-\frac{\beta}{2}\frac{\left(\sum_{j\ne j'}m_jy_{0j}+m_{j'}y_{j'}-\theta\right)^2}{\sum_{j\ne j'}\lambda_j^{-1}m_j^2}}
\end{equation}
Then we integrate over $y_{j'}$, which is again a Gaussian integral with a linear term,
\begin{align}
 \int I(y_{j'})\int I(k)\,dkdy_{j'} &= \left(\frac{2\pi\beta}{\sum_{j\ne j'}\lambda_j^{-1}m_j^2}\right)^{1/2} \int e^{-\frac{\beta}{2}\lambda_{j'}y_{j'}^2 + \beta\lambda_{j'}y_{0j'}y_{j'} -\frac{\beta}{2}\frac{\left(\sum_{j\ne j'}m_jy_{0j}+m_{j'}y_{j'}-\theta\right)^2}{\sum_{j\ne j'}\lambda_j^{-1}m_j^2}}\,dy_{j'}\\
 &= \left(\frac{2\pi\beta}{\sum_{j\ne j'}\lambda_j^{-1}m_j^2} \frac{2\pi\lambda_{j'}^{-1}\sum_{j\ne j'} \lambda_j^{-1}m_j^2}{\beta\sum_j\lambda_{j}^{-1}m_j^2}\right)^{1/2} e^{\frac{\beta}{2}\lambda_{j'}y_{0j'}^2} e^{-\frac{\beta}{2}\frac{\left(\theta - \sum_j m_j y_{0j}\right)^2}{\sum_j \lambda_j^{-1}m_j^2}}\\
 &= 2\pi \left(\lambda_{j'}n^TA^{-1}n\right)^{-1/2} e^{\frac{\beta}{2}\lambda_{j'}y_{0j'}^2} e^{-\beta(\theta-n^Tx_0)^2/2n^TA^{-1}n}
\end{align}
The last equality follows from the fact that $n^TA^{-1}n=m^T\Lambda^{-1}m=\sum_j \lambda_j^{-1}m_j^2$.

Taken together, we see that
\begin{align}
 I &= e^{-\frac{\beta}{2}\lambda_{j'}y_{0j'}^2}(2\pi/\beta)^{N-1}\left(\prod_{j\ne j'}\lambda_{j}^{-1/2}\right)\left(\lambda_{j'}n^TA^{-1}n\right)^{-1/2} e^{\frac{\beta}{2}\lambda_{j'}y_{0j'}^2} e^{-\beta(\theta-n^Tx_0)^2/2n^TA^{-1}n}\\
 &= (2\pi/\beta)^{N-1}\left(\prod_{j}\lambda_{j}^{-1/2}\right)\left(n^TA^{-1}n\right)^{-1/2} e^{-\beta(\theta-n^Tx_0)^2/2n^TA^{-1}n} \\
 &= \frac{e^{-\beta(\theta-n^Tx_0)^2/2n^TA^{-1}n}}{\sqrt{(2\pi/\beta)^{1-N}(n^TA^{-1}n)\det A}}
\end{align}
as desired.

\subsection{Projection of Langer's result to one dimension}

Let us consider the same $N$-dimensional potential of mean force $W(r)$ as in Section \ref{Langer_theory}. Here, we are interested in further projecting $W(r)$ to a one-dimensional potential of mean force. Let us denote this direction by a vector $n$. Integrating away the degrees of freedom orthogonal to $n$ is equivalent to performing a surface integral of the probability density $\pi(r)$ over a family of planes that are orthogonal to $n$. Let us parameterize this family of planes as $n^Tr=\theta$ for $\theta\in\mathbb R$. After having worked through our derivation of Langer's result, it should be obvious to the reader that the one-dimensional potential of mean force is
\begin{equation}
e^{-\beta F(\theta)}=Z^{-1}\int\delta(\theta-n^Tr)e^{-\beta W(r)}\,dr
\end{equation}
Adapting Kramers' result in \eqref{Kramers_rate_1} to $F(\theta)$, the one-dimensional rate constant is
\begin{equation}
 k(n)= \left(\int_{\theta_A}^{\theta_B}\frac{e^{\beta F(\theta)}}{n^TDn}\,d\theta \int_{-\infty}^{\theta^\ddagger}e^{-\beta F(\theta)}\,d\theta\right)^{-1} \label{Kramers_rate_BS_proj}
\end{equation}
where $\theta_A$ and $\theta_B$ are the positions of free energy minima at the reactant and product well, respectively, and $\theta^\ddagger$ is the position of the saddle point in the one-dimensional projection. The expression $n^TDn$ is the one-dimensional diffusion constant along $n$. To understand this expression, first note that since the diffusion matrix represents a physical property of the system, the representation of $D$ should change with coordinate transformations in such a way as to leave the underlying physics invariant. This makes $D$ a tensor, specifically a second-order contravariant tensor. Under the transformation $r\mapsto \theta$, the tensor $D_{ij}$ transforms correspondingly to a zeroth-order tensor (i.e., a constant) by
\begin{equation}
 \sum_{ij}D_{ij}\frac{\partial\theta}{\partial r_i}\frac{\partial\theta}{\partial r_j}=\sum_{ij}n_iD_{ij}n_j=n^TDn
\end{equation}
as desired.

Using \eqref{gaussian_delta_integral} and the harmonic approximations at the saddle point \eqref{Langer_harmonic_saddle}, the first integral in \eqref{Kramers_rate_BS_proj} evaluates to
\begin{align}
 \int_{\theta_A}^{\theta_B}\frac{e^{\beta F(\theta)}}{n^TDn}\,d\theta &\approx Z\frac{e^{\beta\Delta W^\ddagger}}{n^TDn} \int\left(\int \delta(\theta-n^Tr)e^{-\beta(r-r^\ddagger)^TH^\ddagger(r-r^\ddagger)/2}\,dr\right)^{-1}d\theta\\
 &= Z\frac{e^{\beta\Delta W^\ddagger}}{n^TDn}\sqrt{(2\pi/\beta)^{1-N}(n^TH^{-\ddagger}n)\det H^\ddagger}\sqrt{(2\pi/\beta) |n^TH^{-\ddagger}n|}\\
 &= (2\pi/\beta)^{1-N/2}Ze^{\beta\Delta W^\ddagger}\frac{|n^TH^{-\ddagger}n|}{n^TDn}|\det H^\ddagger|^{1/2}
\end{align}
Note that since $\det H^\ddagger<0$, the use of \eqref{gaussian_delta_integral} requires that $n^TH^{-\ddagger}n<0$; otherwise the integral diverges as we discussed in the previous section.  Similarly, with the harmonic approximation at the reactant well \eqref{Langer_harmonic_reactant}, the second integral in \eqref{Kramers_rate_BS_proj} evaluates to
\begin{align}
 \int_{-\infty}^{\theta^\ddagger}e^{-\beta F(\theta)}\,d\theta &\approx Z^{-1} \int\int \delta(\theta-n^Tr)e^{-\beta(r-r_A)^TH_A(r-r_A)/2}\,drd\theta \\
 &= Z^{-1}\sqrt{\frac{(2\pi/\beta) (n^TH^{-1}_An)}{(2\pi/\beta)^{1-N}(n^TH^{-1}_An)\det H_A}}\\
 &= Z^{-1}(2\pi/\beta)^{N/2}\left(\det H_A\right)^{-1/2}
\end{align}
The application of \eqref{gaussian_delta_integral} here does not impose any further conditions on $n$, since $H_A$ is symmetric positive definite. Together, the rate constant is
\begin{equation}
 k(n)= \frac{\beta}{2\pi}\sqrt{\frac{\det H_A}{|\det H^\ddagger|}}\frac{n^TDn}{|n^TH^{-\ddagger}n|}e^{-\beta\Delta W^\ddagger} \label{BS_rate_constant}
\end{equation}

The one-dimensional rate constant in \eqref{BS_rate_constant} is reminiscent of Langer's result in \eqref{Langer_rate_constant}, except with $\lambda_+$ replaced by $n^TDn/|n^TH^{-\ddagger}n|$. In fact, $n^TDn/|n^TH^{-\ddagger}n| = \lambda_+$ when $n$ is parallel to $v_+$ (this should be clear from the fact that $H^{-\ddagger}v_+=-\lambda_+^{-1}Dv_+$.) Furthermore $k(v_+)$ is the minimum of $k(n)$ (more strictly speaking, any vector proportional to $v_+$ will do). To see why, recall the earlier discussion related to the simultaneous diagonalization of $H^\ddagger$ and $D$ by $V$ in \eqref{gen_eig_diag}. With a change of variable $n= Vm$,
\begin{equation}
 \frac{n^TDn}{|n^TH^{-\ddagger}n|} = \frac{m^TV^{T}DVm}{|m^TV^TH^{-\ddagger}Vm|} = \frac{m^Tm}{|m^T\Lambda^{-1} m|} = \frac{\|m\|^2}{|-\lambda_+^{-1}m_+^2 + \sum_{j\ne+}\lambda_j^{-1} m_j^2|}
\end{equation}
The expression $m^T\Lambda^{-1}m/m^Tm$ is an example of a Rayleigh quotient. In the absence of any constraint, the quotient is bounded between the largest and smallest eigenvalues of $\Lambda^{-1}$. This bound is unfortunately not helpful, since we require that $n^TH^{-\ddagger}n$ (and thus $m^T\Lambda^{-1}m$) be negative. For any fixed length $\|m\|$, the expression $|-\lambda_+^{-1}m_+^2 + \sum_{j\ne+}\lambda_j^{-1} m_j^2|$ in the denominator is maximized under this constraint whenever the positive terms in the sum $\sum_{j\ne+}\lambda_j^{-1} m_j^2$ are minimized. This is achieved by setting all $m_j=0$ except for $|m_+| = \|m\|$; in other words, $m$ is proportional to $e_+$, and thus $n$ is proportional to $v_+$ (the change of variable described here is the same as that in Section \ref{flux_surface_integral}, but one should be careful not to confuse the vectors $n$ and $m$ defined in this section with those defined in Section \ref{flux_surface_integral}). Taken together, we have shown that $n=v_+$ minimizes \eqref{BS_rate_constant} and the minimum is equivalent to \eqref{Langer_rate_constant}, as desired.

\section{Acknowledgement}

I would like to thank Robert Alberstein for his critical reading of the note, and I would like to acknowledge support through Chan Zuckerberg Biohub Investigator funds to Dr. Tanja Kortemme (UCSF).

\newpage
\singlespace
\bibliography{zotero_library}
\bibliographystyle{pnas-new.bst}

\end{document}